\documentclass[letter]{IEEEtran}

\usepackage{mathtools}
\usepackage{graphicx,color}
\usepackage{cite}
\usepackage{setspace} 
\usepackage{amsmath,amsthm,amssymb,bm}
\usepackage{multirow}
\usepackage{hhline}
\usepackage{epsfig}
\usepackage{epstopdf}
\usepackage{verbatim}
\usepackage{algorithm}
\usepackage{algpseudocode}
\usepackage{cases}
\usepackage{array}
\usepackage{subfigure}
\usepackage{stfloats}
\usepackage{tabstackengine}
\setstackEOL{\cr}

\newtheorem{definition}{Definition}
\newtheorem{lemma}{Lemma}

\newtheorem{corollary}{Corollary}

\allowdisplaybreaks

\usepackage{forloop}
\newcounter{ct}

\makeatletter 
\def\@eqnnum{{\normalsize \normalcolor (\theequation)}} 
\makeatother

\makeatletter
\renewcommand*\env@matrix[1][\arraystretch]{%
	\edef\arraystretch{#1}%
	\hskip -\arraycolsep
	\let\@ifnextchar\new@ifnextchar
	\array{*\c@MaxMatrixCols c}}
\makeatother
\begin{document}
	
	\title{A General Approach to Fully Linearize the Power Amplifiers in mMIMO with Less Complexity}
	\author{Ganesh Prasad, \textit{Member, IEEE}, Håkan Johansson \textit{Senior Member, IEEE}, and Rabul Hussain Laskar \thanks{G. Prasad and H. Johansson are with the Division of Communication Systems, Department of Electrical Engineering, Linköping University, 581 83 Linköping, Sweden (e-mail: \{ganesh.prasad, hakan.johansson\}@liu.se).
			
	R. H. Laskar is with the Department of Electronics and Communication Engineering, National Institute of Technology Silchar, 788 010 Silchar, India (e-mail: rhlaskar@ece.nits.ac.in).
	}}

	\maketitle
	\begin{abstract}
		A radio frequency (RF) power amplifier (PA) plays an important role to amplify the message signal at higher power to transmit it to a distant receiver. Due to a typical nonlinear behavior of the PA at high power transmission, a digital predistortion (DPD), exploiting the preinversion of the nonlinearity, is used to linearize the PA. However, in a massive MIMO (mMIMO) transmitter,  a single DPD is not sufficient to fully linearize the hundreds of PAs. Further, for the full linearization, assigning a separate DPD to each PA is complex and not economical. In this work, we address these challenges via the proposed low-complexity DPD (LC-DPD) scheme. Initially, we describe the fully-featured DPD (FF-DPD) scheme to linearize the multiple PAs and examine its complexity. Thereafter, using it, we derive the LC-DPD scheme that can adaptively linearize the PAs as per the requirement. The coefficients in the two schemes are learned using the algorithms that adopt indirect learning architecture based recursive prediction error method (ILA-RPEM) due to its adaptive and free from matrix inversion operations. Furthermore, for the LC-DPD structure, we have proposed three algorithms based on correlation of its common coefficients with the distinct coefficients. Lastly, the performance of the algorithms are quantified using the obtained numerical results.
	\end{abstract}   
	\begin{IEEEkeywords}
		Digital predistortion, massive MIMO, direct learning architecture, indirect learning architecture, recursive prediction error method.
	\end{IEEEkeywords}
	
	\section{Introduction}\label{sec:intro}
	In the wireless transmitters, the radio frequency (RF) power amplifiers (PAs) are used to amplify the modulated signals for distant transmissions. However, the in-band and out-of-band nonlinear distortions occur to the signals amplified near to saturation region of the PAs~\cite{ken01}. This can be reduced by employing some backoff to the peak power of the signals. But, it reduces the efficiency of the PAs. Therefore, the preprocessing like digital predistortion (DPD) over the transmit signals before the PAs are required to linearize the resultant signals towards the saturation region. Since a decade, many works have focused on the linearization of multiple power amplifiers in the transmitters like massive MIMO (mMIMO) transmitters. But, they have focused on the linearization in a particular direction of beamforming instead of linearizing all the PAs. Because, the linearization of each PA requires separate DPD block along with the driving RF chain. Thus, due to high complexity, it is not suitable for an economical mMIMO transmitter. To deal with it, in this work, we have proposed a most general approach to fully linearize all the PAs with less complexity. Also, we have discoursed in detail about the fundamentals behind the challenges and the procedure to tackle it. 
	
	\subsection{Related Works}
	The preprocessing using DPD has an inverse property to the nonlinear PA to mitigate the nonlinearties in the desired transmit signal~\cite{kat22}. From the state-of-the-art, mostly the linear parametric models have been used for the the DPD~\cite{mor04}. One of the methods to identify the DPD coefficients is least square (LS) due to its fast convergence~\cite{d21,liu17,wan18}. But, despite mathematical simplicity, its computational complexity is high due to engagement of inverse operations of the matrices of large sizes that correspond to the estimation of large number of DPD coefficients. However, many works have proposed the algorithms to reduce the complexity for the identification of the DPD based on LS method~\cite{zha24,wan23,gua19,gil20}. For example, the size of the matrix is reduced by normalization of the DPD basis functions (BFs) followed by their pruning~\cite{zha24}. Also, based on stationary random process, the time varying matrix associated with the DPD coefficients is replaced by a constant covariance matrix~\cite{wan23}. Further, in an iterative algorithm based on LS, the samples of the DPD coefficients (or the size of the matrix) can be reduced by considering the correlation in the observation errors between two iterations~\cite{gua19}. Besides, the matrix size can also be reduced using the eigenvalue decomposition and principal component analysis (PCA) that decreases the order of the memory polynomial model of the DPD~\cite{bra25,gil20}. In eigenvalue decomposition, the number of DPD coefficients can be reduced by considering only the dominant eigenvectors. Whereas, in PCA, the reduction is achieved by converting the correlated BFs of the DPD into uncorrelated BFs. 
	
	Although, the above techniques help in reduction of the size of the matrices, but, for time varying and highly nonlinear PAs, still, the required number of DPD coefficients is large. Thus, it leads to an undesirable large matrix operations. Therefore, the recursive based algorithms like least mean square (LMS)~\cite{swa31}, recursive least squares (RLS)~\cite{sur28,mor29}, and recursive prediction error method (RPEM)~\cite{gan32} are computationally more reliable at the cost of their slow convergence to the desired optimal value of the variables. Using LMS, the DPD adjusts its coefficients to minimize the mean square error (MSE) between the PA output and the desired signal. The coefficients are updated using stochastic gradient decent method that minimizes the instantaneous error in each iteration. However, LMS is quite unstable and it is very sensitive in the step size for the update~\cite{moh30}. In conventional LS estimation, a batch of input and output data samples of the PA are used to update the DPD coefficients. But, in RLS, using a set of equations, the LS estimation is represented recursively and the coefficients are updated accordingly for the obtained new data sample of the input and output. To discount the influence of older samples, it uses an exponential weighing known as forgetting factor. The chosen value of the forgetting factor gives a trade-off between the precision and convergence and its low value provides high fluctuation to noise. Therefore, the forgetting factor is improved further in RPEM by considering its variation with time~\cite{sod34}. In the existing works, mostly, these adaptive algorithms are applied to two types of DPD learning architectures: (i) direct learning architecture (DLA)~\cite{zho26,paa27} and (ii) indirect learning architecture (ILA)~\cite{sur28,mor29}. DLA has better performance in the presence of noise at the output the PA, but, ILA is more effective in the convergence rate~\cite{cha33}. Therefore, ILA is widely used for the identification of the DPD. Also, in our proposed work, we have considered RPEM algorithm in an ILA architecture for the DPD identification. Next, we describe the state-of-the-art for the linearization of the multi-antenna transmitters. 
	
	In the multi-antenna systems like MIMO or mMIMO, although the PAs in the transmitters are of same type, but, in practice, they have different nonlinearties due to their sensitivity to process, supply voltage, and temperature (PVT)~\cite{cha35,jar36}. Therefore, a single DPD is not capable to linearize all the PAs~\cite{ng02} and ideally, each PA requires a separate DPD~\cite{hau08}. But, the ideal case provides undesirably high complexity in hardware implementation as well as in processing and even not feasible for a mMIMO transmitter where hundreds of PAs need to be linearized. Subsequently, instead of linerizing all PAs, a resultant single PA can be linearized whose output is the sum of the outputs of the PAs~\cite{cho38,luo39}. However, it addresses the average nonlinerites of the PAs, thus, none of them is fully linearized. On the other hand, instead of sum, the beam-oriented (BO) output of the PAs in a given direction can be linearized using a single DPD~\cite{liu08,tar40,liu41}. As it addresses the nonlinearity of the BO output in the desired direction (main lobe), again, the PAs are not fully linearized. Thereby, they are not able to linearize the outputs in other directions that gives the nonlinear sidelobes and typically, their power level is only $10$ dB lesser than the linear main lobe~\cite{yu37}. This can be improved by frequently updating the DPD for different directions. Also, the number of DPD coefficients per update can be reduced using the pruning algorithms~\cite{bri42}. However, the frequent update is not reliable for online operations and it leads to high computational complexity. Moreover, at a time, the DPD is identified for a particular BO direction and the PAs are not fully linearized that still gives the comparable nonlinear sidelobes to the main lobe. If we assume the similar distribution of nonlinearites over the PAs, the side lobes can be reduced by optimally adjusting the amplitude of the phase shifters in the BO output~\cite{ter43}. However, in general, it cannot provide the full linearization of the PAs.  The performance towards the full linearization can be improved by including extra tuning box to each PA. The tuning boxes compensate the nonlinear differences between the PAs such that the resultant nonlinearity is same for each PA, thereafter, using a single DPD, the PAs are fully linearized~\cite{yu44,dia45,zha46}. Nonetheless, for the compensation of the differences, each tuning box is modeled using a polynomial model which needs to be identified using a learning algorithm. Therefore, its complexity is approximately similar to incorporate separate DPD for each PA. Different from only DPD operations, the sidelobes in the BO output can be linearized more reliably using the two layer of operations: the DPD training followed by the post-weighting coefficients optimization that are multipled by the DPD output signal and distributed at the respective PAs' input~\cite{yan07,pra14}. In a simplified analysis, different post-weighting coefficients are assigned for each PA, but, in a branch of a PA, same post-weighting coefficient are multiplied to the BFs of the DPD. Thus, due to less degree of freedom per branch, it is less reliable in post-weighting linearization of the PAs~\cite{yan07}. Also, to distribute different signals to the branches of the PAs, separate RF-chain is needed for each PA that gives a high complexity in a mMIMO transmitter. Later, in our proposed work~\cite{pra14}, we adopted an adaptive post-weighting architecture that increases the degree of freedom (DOF) per branch as well as reduces the number of RF-chains requirement. But, still, due to optimization of post-weighting coefficients for discrete range of directions, the PAs are not fully linearized. 
	
	\subsection{Motivation and Key Contribution}
	As described earlier, the PAs in a multi-antenna transmitter can be fully linearized using identification of a separate DPD to each PA~\cite{hau08}. But, it leads to high complexity in the structure and in the computation to learn the coefficients. Also, for the distribution of the predistorted signals, it requires a separate RF-chain to each PA. Based on it, we propose a most general approach using a low-complexity DPD (LC-DPD) structure which approximates seperate DPD identification requirement as well as the reduction of the number of RF-chains as per the requirement in the mMIMO transmitters. The key contribution of this work is four-fold as follows. (i) First, we deduce the reduction in the number of coefficients for a given type of PAs in a subarray from the measurement data and the obtained numerical result of a system setting. Then, we propose a fully-featured DPD (FF-DPD) scheme to fully linearize the PAs in a subarray and describe its complexity in terms of number of multipliers, adders, and RF chains. (ii) Using the FF-DPD structure, we derive the less complex and non-trivial LC-DPD structure. The number of coefficients in it is reduced based on a geometric sequence and corresponding coefficients are represented in a block vector form. Based on the geometric sequence, we derive the expression of the number of multipliers, adders, and RF chains which are significantly reduced; thus, reduces the complexity. (iii) Next, for the training of the coefficients for the two schemes, we propose four algorithms based on indirect learning architecture based recursive prediction error method (ILA-RPEM): one for the FF-DPD scheme and three for the LC-DPD. Apart from the structural complexity of the FF-DPD, we also describe the computational complexity of its training. The performance of the three algorithms for the LC-DPD is determined based on the correlation of its common coefficients to the distinct coefficients in the structure. It is also shown that the complexities of the three algorithms are less than the algorithm for FF-DPD. Further, for various operations in the four algorithms, we define the four operators and describe their properties. (iv) Lastly, we obtain the numerical results in terms of power spectral density (PSD) and error vector magnitude (EVM) using the algorithms for the two schemes and obtain the various insights by comparing their performances.
	
	\section{Structures for Full Linearization}
	In this section, first, we describe the ideal structure for the predistortion to fully linearize the multiple PAs. Thereafter, we derive a low-complexity structure that approximates the full linearization.
	
	\begin{figure}[!t] 
		\centering \vspace{-0mm} \includegraphics[width=2.7in]{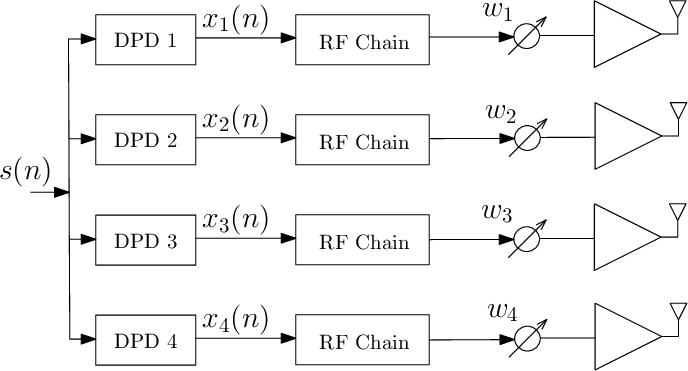}\vspace{-0mm}
		\caption{\small An ideal structure to linearize a subarray of $S=4$ PAs.}    \label{fig:sys_mod_gen}\vspace{-0mm}
	\end{figure}
	If we consider a subarray of $S$ PAs in a mMIMO transmitter as shown in Fig.~\ref{fig:sys_mod_gen} (where $S=4$), ideally, separate DPDs are applied to the respective PAs for the full linearization. As each DPD output signal is different, therefore, a separate RF-chain is employed. Thereafter, the following signals are phase shifted using the analog phase shifters (analog beamforming weights), $\{w_l\}$; $l\in{\{1,\cdots,S\}}$ to get the BO output from the PAs in a specific direction. Based on general memory polynomial (GMP)~\cite{mor04}, the $l$th DPD output $x_l(n)$ to the input message $s(n)$ can be expressed as:
	\begin{align}\label{eq:dpd_out}
		x_{l}(n) = \sum_{p=0}^{P_l-1}\sum_{m=0}^{M_l-1}\phi_{p,m}^l s(n-m) |s(n-m)|^p,
	\end{align}
	where $\phi_{p,m}^l$ is the coefficient for the BF, $s(n-m) |s(n-m)|^p$ of $p$th power and $m$th delay. Eq.~\eqref{eq:dpd_out} represents the most general model where the memory length $M_l$ and the order $P_l$ of the polynomial depends on the $l$th PA. The outputs\footnote{For convenience, the time marker index of the signals are omitted.} of the $S$ DPDs in the subarray are represented by a vector $\bm{X}=[x_1,\cdots,x_S]^T$. Further, $x_l$ is multiplied by the beamforming weight $w_l$ and inputted to the respective $l$th nonlinear PA. Output $y_l(n)$ of the PA can be expressed as:
	\begin{align}\label{eq:out_pa}
		y_l(n) = f_{non}^l(w_lx_l(n)),
	\end{align}
	where $f_{non}^l(\cdot)$ represents the nonlinear function for the $l$th PA. For the $S$ PAs in the subarray, the output vector $\bm{Y}$ can be expressed as: $\bm{Y} = [y_1,\cdots,y_S]^T$. Nevertheless, the implementation of the general architecture in Fig.~\ref{fig:sys_mod_gen} to completely linearize all the PAs is highly complex. Because, the different set of BFs with their coefficients for each of the DPDs require many delays, multipliers, and adders. Further, the computational complexity of the iterative/learning algoirthm to identify the coefficients for each of the DPDs is undesirably high. Also, the number of RF chains is same as the number of PAs, $S$ in the subarray which is not economical for a mMIMO transmitter. 
	
	\begin{figure}[!t] 
		\centering \vspace{-0mm} \includegraphics[width=2.7in]{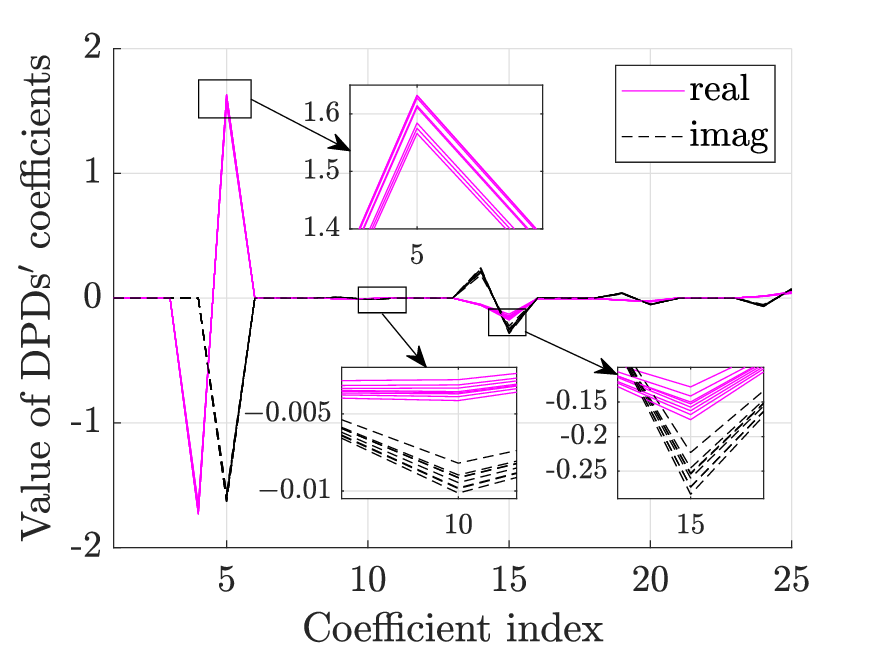}\vspace{-0mm}
		\caption{\small Real and imaginary values of identified $8$ DPDs' coefficients to predistort respective $8$ PAs based on Saleh model~\cite{sal47}.}    \label{fig:coeff_val}\vspace{-0mm}
	\end{figure}
	
	In order to simplify the structure, first, we analyze the values of the identified DPDs' coefficients for the given PAs in a subarray. In Fig.~\ref{fig:coeff_val}, we have plotted the real and imaginary values of the identified coefficients of $8$ DPDs to fully linearize the respective $S=8$ traveling-wave tube (TWT) PAs based on Saleh model~\cite{sal47} having different AM/AM and AM/PM nonlinearities as described in Section~\ref{sec:num_rel}. Here, $P_l=5$ and $M_l =5$; $\forall l\in\{1, \cdots, S\}$. The DPDs are trained using the adaptive ILA-RPEM algorithm which is described in detail in the following sections. From the figure, it can be observed that out of $25$ BFs, the coefficients of some BFs with indices in the set $\mathcal{I}=\{4,5,9,10,14,15,19,20,24,25\}$ are non-zero. Further, the index set $\mathcal{I}$ of the BFs of non-zero coefficients is same for all PAs, because, the PAs are of same type\footnote{Note that in the supplementary file of~\cite{bri10}, from the measurement of outputs of 16 HMC943APM5E PA ICs for the input signal at $28.5$ GHz, all the PAs nonlinerties are identified using the coefficients of same BFs. Therefore, they can be linearized using the DPDs' coefficients of same BFs.}. Also, the deviation in the values of a coefficient for different PAs is higher for higher value of the coefficient than the coefficients of lower values. For example, the mean deviations for the indices $5$ and $10$ are $0.0249$ and $4.0248\times 10^{-4}$ for real part and $0.0248$ and $5.7523\times 10^{-4}$ for the imaginary part, respectively. Thus, the coefficients with higher values dominate in the linarization of the $S$ PAs. Based on these observations, next, we reduce the number of coefficients in the proposed two DPD schemes. 
	 

\begin{figure}[!t]
	\centering
	\subfigure[]{\includegraphics[width=3.5in]{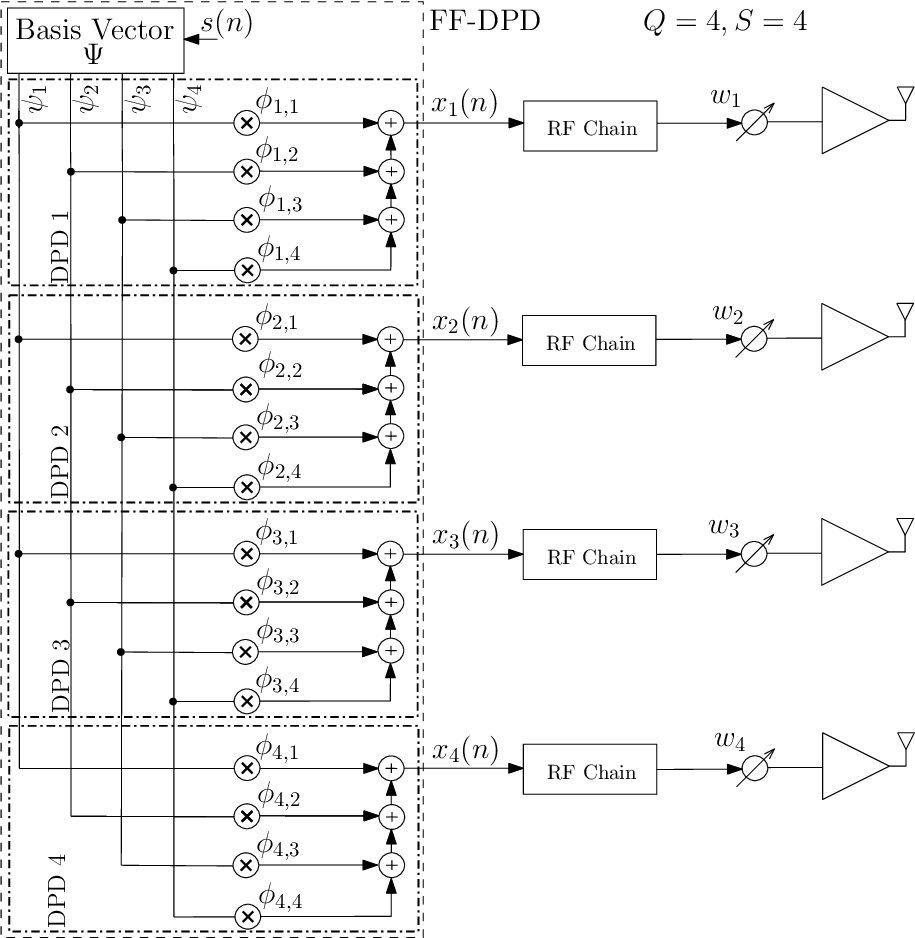}}
	\subfigure[]{\includegraphics[width=3.5in]{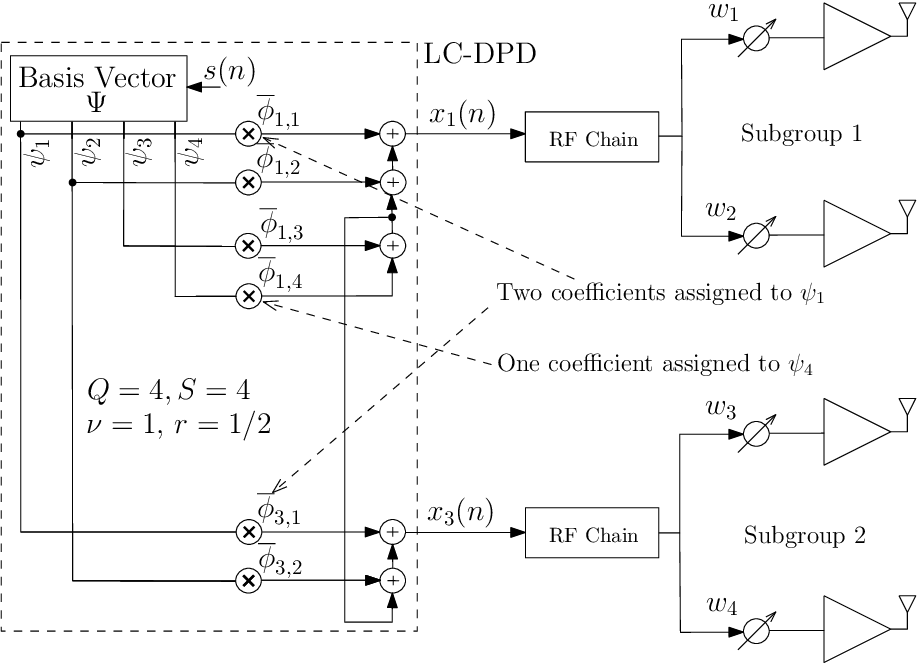}}
	\caption{\small Performance comparison of the proposed PW schemes against (a) the benchmark schemes and (b) the systems with only BO-DPD.}
	\label{fig:sys_mod}\vspace{-0mm}
\end{figure}
	
\subsection{Fully-Featured DPD (FF-DPD)}\label{sec:FF-DPD}
As described earlier in this section, the predistortion signals for a given type of PAs can be obtained using the same set of BFs based on GMP (cf. Fig.~\ref{fig:coeff_val}). Thus, in FF-DPD, we consider the same set of BFs of order $P$ and memory length $M$ for all the PAs in the subarray. Therefore, the $l$th output $x_l$ from the FF-DPD can be obtained again using~\eqref{eq:dpd_out} after substituting $P_l=P$ and $M_l=M$. The total number of BFs in the set $\{s(n-m) |s(n-m)|^p\}$ for $p\in\{0,\cdots,P-1\}$ and $m\in\{0,\cdots,M-1\}$ is $PM$. However, as observed in Fig.~\ref{fig:coeff_val}, out of $PM$ BFs, some BFs have nonzero DPD coefficients for a given type of PAs. Therefore, in general, we represent $Q$ BFs as a vector $\bm{\Psi}=[\psi_1,\cdots,\psi_Q]^T$ with their respective nonzero coefficient vector for the $l$th PA as $\bm{\Phi}_l =[\phi_{l,1},\cdots,\phi_{l,Q}]^T$, where $\psi_i$ is the $i$th BF\footnote{BF $\psi_i$ is a function of $s(n)$, is given by $\psi_i(s(n))$ for $i\in\{1,\cdots,Q\}$.} and $\phi_{l,i}$ is its nonzero coefficient for $i\in\{1,\cdots,Q\}$. Using $\bm{\Psi}$ and $\bm{\Phi}_l$, $x_l$ in~\eqref{eq:dpd_out} can be expressed in matrix form as:
\begin{align}\label{eq:x_l_FF}
x_l = \bm{\Phi}_l^T\bm{\Psi},
\end{align}
Using $x_l$, the output $y_l$ of the $l$th PA is obtained using the same process as in~\eqref{eq:out_pa} where $x_l$ is multiplied by the phase shifter $w_l$ and inputted to the PA to get $y_l$, thus, output vector $\bm{Y}$ is obtained. Moreover, the coefficient vectors for the $S$ PAs in the subarray can be expressed in a block vector as: $\bm{\Phi}=[\bm{\Phi}_1^T,\cdots,\bm{\Phi}_S^T]^T$. From Fig.~\ref{fig:sys_mod}(a), $S$ coefficients are multiplied by each BF. Thus, for $Q$ BFs, the FF-DPD structure has $QS$ coefficients. Besides, the number of multipliers, $N_m^F$ in the structure is same as the number of coefficients, i.e., $N_m^F=QS$. Further, using the structure of FF-DPD in Fig.~\ref{fig:sys_mod}(a), the number of adders, $N_a^F$ can be determined as follows\footnote{In this work, the numbers of adders for the different structures are determined for the assumption that an adder has two inputs and one output.}. Each predistorted output ($x_l$; $l\in\{1,2,3,4\}$) is determined though the sum of the multiplications of the $Q$ coefficients by respective BFs. Therefore, in generation of each output, the number of adders is $(Q-1)$, i.e., one less than the number of coefficients. Thus, for $S$ outputs, the total number of adders, $N_a^F=(Q-1)S$. Moreover, the number of RF chains, $N_{RF}^F$ is same as the number of PAs, i.e., $N_{RF}^F=S$. For instance, Fig.~\ref{fig:sys_mod}(a) depicts the FF-DPD for $S=4$ and $Q=4$. It has $N_m^F=16$, $N_a^F=12$, and  $N_{RF}^F=4$. If we compare Fig.~\ref{fig:sys_mod}(a) with Fig.~\ref{fig:sys_mod_gen}, the ideal structure for the linearization of the subarray is same as the structure of FF-DPD except the same set of BFs has been used for all PAs in the later. Thus, the complexity of FF-DPD is still high in terms of multipliers, adders, and the RF chains. These complexities can be reduced using LC-DPD which is described next.

\subsection{Low-Complexity DPD (LC-DPD)}\label{sec:LC-DPD}
Using LC-DPD, we reduce the complexity to fully linearize the PAs as follows. As described earlier (cf. Fig.~\ref{fig:coeff_val}), coefficients of a dominant BF in the generation of the predistorted signals for different PAs has higher deviation in its value. Therefore, to reduce the number of coefficients in LC-DPD, the BFs in the vector $\bm{\Psi}=[\psi_1,\cdots,\psi_Q]^T$ are arranged in decreasing order of their dominance. Then, more coefficients are multiplied by a higher dominant BFs than the lower dominant BFs to generate the predistorted signals. Therefore, different from FF-DPD, in LC-DPD, the number of coefficients are multiplied adaptively by the BFs according to their dominance. Therefore, we decrease the number of coefficients based on a geometric series as follows. 
\begin{eqnarray}\label{eq:seq_grp}
\left\{\hspace{-0mm}\begin{array}{cc}
\hspace{-0mm}\{n_1Sr^\nu,\cdots, n_gSr^{(\nu+g-1)}\},&  \!\!\!\!\text{ Case I} \\ 
\{n_1Sr^\nu,\cdots, n_{g-1}Sr^{(\nu+g-2)},n_{g}\times 1\},&  \text{ Case II}
\end{array}\hspace{-2mm}\right.
\end{eqnarray}
where $\nu\in\mathbb{Z}^+=\{0,1,\cdots\}$, $\sum_{i=1}^{g}n_i=Q$; $n_i\in \mathbb{P}=\{1,2,\cdots\}$ and the cases are: Case I: $S r^{(\nu+q-1)}\ge 1$ and Case II: $\{S r^{(\nu+g-2)}\ge 1\}\wedge\{S r^{(\nu+g-1)}< 1\}$. According to the sequence in~\eqref{eq:seq_grp}, the $Q$ BFs are divided into $g$ groups, where each of the $n_i$ BFs in the $i$th group are multiplied by $Sr^{(\nu+i-1)}$ coefficients; thus, the total coefficients in the $i$th group is $n_iSr^{(\nu+i-1)}$. Further, over the groups, the number multiplied coefficients per BF decreases as a geometric sequence with a common ratio $r$ $(<1)$. The sequence in Case~II is same as in Case~I except each BF in the last group $g$ is multiplied by one coefficient in Case~II, because, $S r^{(\nu+g-1)}< 1$ and the number of coefficient per BF cannot be a fraction. Now, using this sequence, we define the coefficient vector $\bm{\overline{\Phi}}$ of the LC-DPD as below.
\begin{definition}
	For the sequence of the number of coefficients multiplied by different BFs as expressed in~\eqref{eq:seq_grp}, the coefficient vector $\bm{\overline{\Phi}}$ for the LC-DPD can be represented as:
		\begin{subequations}\label{eq:block_vec}
		\begin{align}\label{eq:block_vec_a}
		\hspace{-1.725in}\bm{\overline{\Phi}} = \left[\bm{\overline{\Phi}}_1^T,\cdots,\bm{\overline{\Phi}}_g^T\right]^T, \hspace{10.5mm}
		\end{align}
		\vspace{-4mm}
		\begin{figure*}[!t]
			\begin{align}\label{eq:Phi_j}\nonumber
			\bm{\overline{\Phi}}_i\hspace{-1mm} =&\hspace{-1mm} \left[\hspace{-0.2mm}\phi_{_{1,\left(\sigma_{i}\hspace{-0.2mm}+\hspace{-0.2mm}1\right)}}\hspace{-0.2mm},\hspace{-0.2mm} \phi_{_{\left(\hspace{-0.2mm}1\hspace{-0.2mm}+\hspace{-0.2mm}r^{-(\hspace{-0.2mm}\nu\hspace{-0.2mm}+\hspace{-0.2mm}i\hspace{-0.2mm}-\hspace{-0.2mm}1\hspace{-0.2mm})}\hspace{-0.2mm}\right),\left(\hspace{-0.2mm}\sigma_{i}\hspace{-0.2mm}+\hspace{-0.2mm}1\hspace{-0.2mm}\right)}}\hspace{-0.2mm}, \hspace{-0.2mm}\cdots\hspace{-0.2mm},\hspace{-0.2mm} \phi_{_{\left(\hspace{-0.2mm}1\hspace{-0.2mm}+\hspace{-0.2mm}\left(\hspace{-0.2mm}Sr^{\left(\hspace{-0.2mm}\nu\hspace{-0.2mm}+\hspace{-0.2mm}i\hspace{-0.2mm}-\hspace{-0.2mm}1\hspace{-0.2mm}\right)}\hspace{-0.2mm}-\hspace{-0.2mm}1\hspace{-0.2mm}\right)r^{-(\hspace{-0.2mm}\nu\hspace{-0.2mm}+\hspace{-0.2mm}i\hspace{-0.2mm}-\hspace{-0.2mm}1\hspace{-0.2mm})}\hspace{-0.2mm}\right),\left(\hspace{-0.2mm}\sigma_{i}\hspace{-0.2mm}+\hspace{-0.2mm}1\hspace{-0.2mm}\right)}}\hspace{-0.2mm},\hspace{-0.2mm}\cdots\hspace{-0.2mm},\hspace{-0.2mm}\phi_{_{1,\left(\hspace{-0.2mm}\sigma_{i}\hspace{-0.2mm}+\hspace{-0.2mm}n_i\hspace{-0.2mm}\right)}}\hspace{-0.2mm},\hspace{-0.2mm} \phi_{_{\left(1+r^{-(\nu+i-1)}\right),\left(\hspace{-0.2mm}\sigma_{i}\hspace{-0.2mm}+\hspace{-0.2mm}n_i\hspace{-0.2mm}\right)}}\hspace{-0.2mm},\hspace{-0.2mm} \cdots\hspace{-0.2mm},\right.\\
			&\left. \hspace{-0.2mm}\phi_{_{\left(\hspace{-0.2mm}1\hspace{-0.2mm}+\hspace{-0.2mm}\left(\hspace{-0.2mm}Sr^{(\hspace{-0.2mm}\nu\hspace{-0.2mm}+\hspace{-0.2mm}i\hspace{-0.2mm}-\hspace{-0.2mm}1\hspace{-0.2mm})}\hspace{-0.2mm}-\hspace{-0.2mm}1\hspace{-0.2mm}\right)r^{-(\hspace{-0.2mm}\nu\hspace{-0.2mm}+\hspace{-0.2mm}i\hspace{-0.2mm}-\hspace{-0.2mm}1\hspace{-0.2mm})}\hspace{-0.2mm}\right),\left(\hspace{-0.2mm}\sigma_{i}\hspace{-0.2mm}+\hspace{-0.2mm}n_i\hspace{-0.2mm}\right)}}\hspace{-0.2mm}\right]^T
			\end{align}
			\hrulefill
		\end{figure*}
	     \begin{align}\label{eq:Phi_q}
	     	\hspace{-2mm}\bm{\overline{\Phi}}_g\hspace{-0.3mm} = \hspace{-0.3mm}\left[\hspace{-0.3mm}\phi_{_{1,\left(\sigma_{g}\hspace{-0.3mm}+\hspace{-0.3mm}1\right)}}, \phi_{_{1,\left(\sigma_{g}\hspace{-0.3mm}+\hspace{-0.3mm}2\right)}},\cdots,\phi_{_{1,\left(\sigma_{g}\hspace{-0.3mm}+\hspace{-0.3mm}n_g\right)}}\hspace{-0.3mm}\right]^T; \text{ for Case II}
	     \end{align}
	\end{subequations}
\noindent	where $\bm{\overline{\Phi}}_i$ is given in~\eqref{eq:Phi_j} for $i\in\{1,\cdots,g\}$ for both the cases except $\bm{\overline{\Phi}}_g$ for Case II is obtained by~\eqref{eq:Phi_q}. Besides, $\sigma_{i}=\sum_{j=1}^{i-1}n_j$ and $\phi_{_{l,q}}$ is the coefficient which is multiplied by the $q$th BF $\psi_q$ to contribute to the $l$th predistorted output of the LC-DPD.
\end{definition}
\begin{IEEEexample}
		For LC-DPD in Fig.~\ref{fig:sys_mod}(b) where $S=4$ and $Q=4$, the BFs in the vector $\bm{\Psi}=[\underbrace{\psi_1,\psi_2}_{n_1=2},\underbrace{\psi_3,\psi_4}_{n_2=2}]^T$ are divided into $g=2$ groups and each group contains two BFs. For $\nu=1$ and $r=1/2$, the sequence of number of coefficients is: $\{n_1Sr^\nu, n_2Sr^{\nu+1}\}=\{4,2\}$; thus, the total number of coefficients is $4+2=6$. Here, each of the BF in the first group is assigned by $Sr^\nu = 2$ coefficients, while for the second group, it is $Sr^{\nu+1}=1$ as depicted in the figure. Further, for the LC-DPD coefficient vector $\bm{\overline{\Phi}}$, the Case I in~\eqref{eq:block_vec_a} is satisfied, thus, using~\eqref{eq:block_vec}, $\sigma_1=0$, and $\sigma_2=2$, the coefficient vector $\bm{\overline{\Phi}}=[\underbrace{\phi_{1,1},\phi_{3,1},\phi_{1,2},\phi_{3,2}}_{\text{elements of }\bm{\overline{\Phi}}_1}, \underbrace{\phi_{1,3},\phi_{1,4}}_{\text{elements of }\bm{\overline{\Phi}}_2}]$.
\end{IEEEexample}

\noindent Furthermore, the total number of coefficients (or multipliers) in $\bm{\overline{\Phi}}$, the number of  adders, and the number of RF chains in the structure of LC-DPD can be determined using Lemma~\ref{lemma1} as described below.  
	\begin{lemma}\label{lemma1}
		 Using the sequence in~\eqref{eq:seq_grp}, the total number of coefficients (or multipliers), $N_m^L$ in $\bm{\overline{\Phi}}$ can be determined as in~\eqref{eq:N_m_a}. Further, for a special case when $n_i=n_j=n$; $\forall i\ne j$, $N_{m,eq.}^L$ is expressed in~\eqref{eq:N_m_b}. Moreover, the number of adders, $N_a^L$ and the number of RF chains, $N_{RF}^L$ are given by~\eqref{eq:N_a_RF}.
		\begin{subequations}\label{eq:N_m}
			\begin{eqnarray}\label{eq:N_m_a}
				&\hspace{-10.5mm}N_m^L\hspace{-1mm} =\hspace{-1mm}
				\left\{\hspace{-2.5mm}\begin{array}{cc}
					Sr^\nu\sum_{i=1}^{g}n_ir^{i-1}\big|\sum_{i=1}^{g}n_i=Q; \!\!\!\!\!&  \text{ Case I} \\ Sr^\nu\sum_{i=1}^{g-1}n_ir^{i-1}+n_g;\!\!\!\!\!&  \text{ Case II},
				\end{array}\hspace{-2mm}\right.\\\label{eq:N_m_b}
				&\hspace{-7.5mm}N_{m,eq.}^L\hspace{-1mm} =\hspace{-1mm}
				\left\{\hspace{-2.5mm}\begin{array}{cc}
					nSr^\nu(1-r^g)/(1-r); \!\!\!\!\!&  \text{ Case I} \\ nSr^\nu(1-r^{(g-1)})/(1-r)+n_g;\!\!\!\!\!&  \text{ Case II}.
				\end{array}\hspace{-2mm}\right.
			\end{eqnarray}
		\end{subequations}
		\begin{align}\label{eq:N_a_RF}
			\hspace{-0.87in}N_a^L = N_m^L -\left \lceil Sr^{(\nu+g-1)}\right\rceil,\;\; N_{RF}^L = Sr^\nu
		\end{align}
	\end{lemma}
	\begin{IEEEproof}
		The expression of $N_m^L$ in~\eqref{eq:N_m_a} is trivial which is the sum of the terms of the sequences in~\eqref{eq:seq_grp} for the two cases. After substituting the condition of the special case: $n_i=n_j=n$; $\forall i\ne j$ in~\eqref{eq:N_m_a}, $n$ is taken common, thus, it becomes a geometric series with common ratio $r$. After simplifying, we get $N_{m,eq.}^L$ as in~\eqref{eq:N_m_b}. The number of adders, $N_a^L$ in the LC-DPD structure can be determined as follows. As described in the structure of FF-DPD, for generation of a predistorted output, if the coefficients to the respective BFs are completely different, then, the number of adder used in the generation is $Q-1$, i.e., one less than the number of coefficients. It can also be observed for the output $x_1$ of the LC-DPD in Fig.~\ref{fig:sys_mod}(b). But, for the output $x_3$, the coefficients for BFs $\psi_1$ and $\psi_2$ are different, while the coefficients for $\psi_3$ and $\psi_4$ are equal to the respective coefficients for $x_1$. We find that for the generation of $x_3$, the number of adders is $2$ which is equal to the number of different coefficients. Based on it, the total number of adders, $N_a^L$ is obtained as in~\eqref{eq:N_a_RF} which is equal to the total number of coefficients (or multipliers) subtracted by the number of predistorted outputs which uses completely different set of coefficients, given by: $Sr^{(\nu+g-1)}$. For Case~II in~\eqref{eq:seq_grp}, $Sr^{(\nu+g-1)}<1$ and only one output is generated by completed different set of coefficients. So, $\lceil Sr^{(\nu+g-1)}\rceil$ represents the number of different sets of coefficients, thus, $N_a^L$ in~\eqref{eq:N_a_RF} is for both the cases. Moreover, the number of RF chains, $N_{RF}^L$ depends on the number of coefficients multiplied by each of the BF in the first group, i.e., $N_{RF}^L=Sr^{\nu}$ as expressed in~\eqref{eq:N_a_RF}. For instance, in Fig.~\ref{fig:sys_mod}(b), each of $\psi_1$ and $\psi_2$ are assigned by $4\times (1/2)^{1}=2$ coefficients, thus, $N_{RF}^L=2$. 
	\end{IEEEproof}
	
	\noindent If we compare the LC-DPD to FF-DPD in Fig.~\ref{fig:sys_mod}, the number of multipliers, adders, and RF chains in LC-DPD are reduced by the factors, $N_m^F/N_m^L=2.67$, $N_a^F/N_a^L=2.4$, and $N_{RF}^F/N_{RF}^L=2$, respectively.

	\section{Training Based on ILA-RPEM: Part I}\label{sec:train_RPEM}
	Now, we describe the ILA-RPEM based learning for FF-DPD and LC-DPD schemes to fully linearize a subarray of PAs. In Part I, we focus on the learning using FF-DPD structure where  first, we describe the learning for FF-DPD coefficients, then, the learning for LC-DPD coefficients is realized by utilizing the structure of FF-DPD. Whereas, in Part II, the learning completely exploits the structure of LC-DPD. 
	
	\begin{figure}[!t] 
		\centering \vspace{-0mm} \includegraphics[width=2.4in]{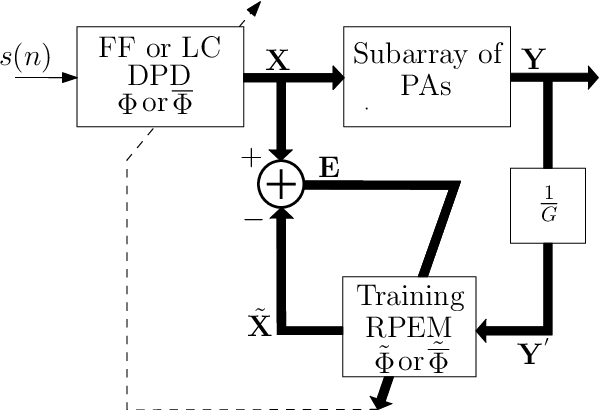}\vspace{-0mm}
		\caption{\small Indirect learning architecture (ILA) based on RPEM.}    \label{fig:train_dpd}\vspace{-0mm}
	\end{figure}

    Fig.~\ref{fig:train_dpd} represents a general ILA architecture to linearize a subarray of PAs using RPEM algorithm.  Here, the message $s(n)$ is inputted to the FF-DPD (or LC-DPD) block with coefficient vector $\bm{\Phi}$ (or $\bm{\overline{\Phi}}$) which generates a vector $\bm{X}$ of predistorted signals using \eqref{eq:x_l_FF}. Thereafter, these signals are inputted to the respective PAs to get the output vector $\bm{Y}$ using \eqref{eq:out_pa}. To minimize the nonlinearties in $\bm{Y}$, in the feedback loop, first, $\bm{Y}$ is scaled by the inverse gain $1/G$ of the PAs to get $\bm{Y}^{'}=[y_1^{'}(n) \cdots y_S^{'}(n)]^T$, where $y_l^{'}(n)=(1/G)y_l(n)$. Then, it is inputted to the training block. Based on RPEM, the block estimates the FF-DPD (or LC-DPD) coefficient vector $\bm{\tilde{\Phi}}$ (or $\bm{\tilde{\overline{\Phi}}}$), where $\bm{\tilde{\Phi}}$ is defined similar to $\bm{\Phi}$ as in Section~\ref{sec:FF-DPD} for FF-DPD structure, whereas, $\bm{\tilde{\overline{\Phi}}}$ is defined as $\bm{\overline{\Phi}}$ in~\eqref{eq:block_vec} for LC-DPD structure. Thereafter, it is copied to FF-DPD (or LC-DPD) block, i.e., $\bm{\Phi}=\bm{\tilde{\Phi}}$ (or $\bm{\overline{\Phi}}=\bm{\tilde{\overline{\Phi}}}$) and again generates $\bm{X}$ followed by $\bm{Y}$. The process repeats until $\bm{\Phi}$ (or $\bm{\overline{\Phi}}$) converges. Thereafter, the FF-DPD (or LC-DPD) is trained to fully linearize the PAs. 

	\subsection{Linearization of a PA using ILA-RPEM}
	We study the processing behind the linearization of $l$th PA of the subarray using the ILA-RPEM in Fig.~\ref{fig:train_dpd} as follows. Using the input $s(n)$, the predistorted output $x_l(n)$ followed by the PA output $y_l(n)$ are obtained using \eqref{eq:x_l_FF} and \eqref{eq:out_pa}, respectively. To capture the inverse of the nonlinear behavior of the PA, the scaled output of the PA $y_l^{'}(n)$ is inputted to the RPEM based learning block which generates the postdistorter signal $\tilde{x}_l(n)$ similar to~\eqref{eq:x_l_FF} as: 
	\begin{align}\label{eq:x_l_tilde}
	\tilde{x}_l(n) = \bm{\tilde{\Phi}}_l^T\bm{\Psi}_l^{'},
	\end{align}
	where $\bm{\tilde{\Phi}}_l$ is the $l$th vector of the block vector $\bm{\tilde{\Phi}}$ and $\bm{\Psi}_l^{'}$ is the vector of BFs defined using same polynomial terms as in $\bm{\Psi}$, only the difference is that instead of $s(n)$, $y_l^{'}(n)$ is the input to the BFs in $\bm{\Psi}$. Thus, the $i$th element of the vector $\bm{\Psi}_l^{'}$ is: $\psi_{l,i}^{'}=\psi_i(y_l^{'}(n))$. Now, the goal of the RPEM algorithm is to minimize the difference between the postdistorted signal $\tilde{x}_l$ in \eqref{eq:x_l_tilde} and the predistorted signal $x_l$ in \eqref{eq:x_l_FF} iteratively by optimizing $\bm{\tilde{\Phi}}_l$. At the end of each iteration, $\bm{\tilde{\Phi}}_l$ is copied to $\bm{\Phi}_l$, i.e., $\bm{\Phi}_l=\bm{\tilde{\Phi}}_l$ and using it, the algorithm tries to capture the inverse of the nonlinear characteristics of the PA. Thus, using the estimated value $\bm{\widehat{\Phi}}_l$ at the end of the convergence, the FF-DPD generates the optimal predistored signal $\widehat{x}_l$ to linearize the $l$th PA. 
	
	Now, we describe the process to minimize the difference $e_l(n) = x_l(n)-\tilde{x}_l(n)$ in each iteration. The cost function $\mathcal{L}(\bm{\tilde{\Phi}}_l)$ based on the average of the power content of $e_l(n)$ over the long horizon is defined as:
	\begin{align}\label{eq:cost_fcn}
	\mathcal{L}(\bm{\tilde{\Phi}}_l) = \textstyle\lim_{N\to\infty} \frac{1}{N}\sum_{n=1}^{N}\mathbb{E}\left[|e_l(n)|^2\right],
	\end{align}
	From~\cite{lju15}, $\mathcal{L}(\bm{\tilde{\Phi}}_l)$ in~\eqref{eq:cost_fcn} can be minimized using the negative gradient of $e_l(n)$ in $\bm{\tilde{\Phi}}_l$. It can be obtained as:
	\begin{align}\label{eq:grad_err}
		-\frac{\mathrm{d}e(n)}{\mathrm{d}\bm{\tilde{\Phi}}_l}=\frac{\mathrm{d}\tilde{x}_l(n)}{\mathrm{d}\bm{\tilde{\Phi}}_l} = \frac{\mathrm{d}\bm{\tilde{\Phi}}_l^T\bm{\Psi}^{'}_l}{\mathrm{d}\bm{\tilde{\Phi}}_l} = \bm{\Psi}^{'}_l.
	\end{align}
	Using the gradient in~\eqref{eq:grad_err}, the training block performs the below computations in~\eqref{eq:rpem_compt} based on RPEM to get the trained coefficients $\bm{\tilde{\Phi}}_l^{(n)}$ for the $l$th PA in the $n$th iteration~\cite{lju15}.
	\begin{subequations}\label{eq:rpem_compt}
		\begin{align}\label{eq:rpem_compt_a}
			\!\!\!\!e_l(n) &= x_l(n)-\tilde{x}_l(n) ,\!\!\!\!\!\!\!\\\label{eq:rpem_compt_b}
			\!\!\!\!\xi_l^{(n)} &=\rho\xi_l^{(n-1)}+1-\rho,\!\!\!\!\!\!\!\\\label{eq:rpem_compt_c}
			\!\!\!\!Z_l^{(n)} &= \bm{\Psi}_l^{'T}(n)P_l^{(n-1)}{\bm{\Psi}_l^{'}}^*(n) + \xi_l^{(n)},\!\!\!\!\!\!\!\\\label{eq:rpem_compt_d}
			\!\!\!\!P_l^{(n)} \hspace{-1mm}&=\hspace{-1mm} (P_l^{(n-1)}\hspace{-2mm} - \hspace{-1mm} P_l^{(n-1)}\hspace{-0.5mm}{\bm{\Psi}_l^{'}}^*(n)\hspace{-0.5mm}{Z_l^{(n)}}^{-1}\hspace{-2mm}\bm{\Psi}_l^{'T}(n)P_l^{(n-1)}\hspace{-0.5mm})/\xi_l^{(n)}, \!\!\!\!\!\!\!\\\label{eq:rpem_compt_e}
			\!\!\!\!\bm{\tilde{\Phi}}_l^{(n)} &=  \bm{\tilde{\Phi}}_l^{(n-1)} + P_l^{(n)}{\bm{\Psi}_l^{'}}^*(n)e_l(n), \!\!\!\!\!\!\!
		\end{align}
	\end{subequations}
	Here, in~\eqref{eq:rpem_compt_a}, $e_l(n)$ is computed. Thereafter, the forgetting factor $\xi_l$ is determined recursively in~\eqref{eq:rpem_compt_b} using its value in the previous iteration and the rate of growth $\rho$. The initial value $\xi_l^{(0)}=\lambda_0$ and $\xi_l$ grows exponentially to $1$ with iterations. Using $\xi_l$, BF vector $\bm{\Psi}_l^{'}$, and the covariance matrix $P_l$, the scalar $Z_l$ is computed in \eqref{eq:rpem_compt_c}. Initial value $P_l^{(0)}=\mu\bm{I}$, where $\bm{I}$ is the identity matrix and $\mu$ is a constant. Finally, matrix $P_l$ is obtained in~\eqref{eq:rpem_compt_d} followed by  $\bm{\tilde{\Phi}}_l$ is determined in~\eqref{eq:rpem_compt_e} recursively at the end of the iteration. Moreover, RPEM is free from complex matrix inverse operations like in a LS estimation as $Z_l$ is scalar. Now, using the above study for the linearization of $l$th PA, we realize the full linearization of PAs of a subarray.
	
	\subsection{FF-DPD to Linearize a Subarray using ILA-RPEM}
	As described earlier, the structure of FF-DPD in Fig.~\ref{fig:sys_mod}(a) is similar to assigning individual DPD in Fig.~\ref{fig:sys_mod_gen} to each of the PAs. Its complexity is reduced only due to utilizing same set of $Q$ BFs for a given type of PAs.  Therefore, using FF-DPD, the full linearization for $S$ PAs through ILA-RPEM is same as the parallel linearization of each of the $S$ PAs using the same process as for the $l$th PA (cf. previous paragraph). For the parallel operations, the different parameters are arranged in the matrix form as follows. We define, $\bm{E}=[e_1,\cdots,e_S]^T$, $\bm{\xi}\triangleq\text{diag}(\xi_1,\cdots,\xi_S)$, $\bm{Z}\triangleq \text{diag}(Z_1,\cdots,Z_S)$, $\bm{\Upsilon}\triangleq \text{diag}(\bm{\Psi}_1^{'},\cdots,\bm{\Psi}_S^{'})$, $\bm{P}\triangleq \text{diag}(P_1,\cdots,P_S)$, and $\bm{\Xi} \triangleq \text{diag}(\xi_1\bm{I}_Q,\cdots,\xi_S\bm{I}_Q)$. Here, $\text{diag}(\cdot)$ is diagonal matrix constructed using the input scalar or matrix elements. 
	\begin{algorithm}[!t]
		{\small
			\caption{\small Iterative estimation of coefficients for FF-DPD.}\label{algo_1}
			\begin{algorithmic}[1]
				\Require The values of $\rho$, $\lambda_0$, $\mu$, and $\bm{\tilde{\Phi}}^{(0)}$
				\Ensure The estimated coefficient vector $\bm{\widehat{\Phi}}$
				\State $\bm{P}^{(0)}=\text{diag}(\underbrace{\mu\bm{I}_Q,\cdots,\mu\bm{I}_Q}_{S \text{ times}})$\label{step1}
				\State $\bm{\xi}^{(0)}=\lambda_0\bm{I}_S$\label{step2}
				\State Assign $\bm{\Phi}^{(0)}=\bm{\tilde{\Phi}}^{(0)}$ and $\bm{X}(1)$ using~\eqref{eq:x_l_FF} followed by determine $\bm{\Upsilon}(1)$ using the outputs $\bm{Y}(1)$ of the PAs\label{step3}
				\State $\bm{\tilde{X}}(1) = \bm{\Upsilon}(1)^T\bm{\tilde{\Phi}}^{(0)}$\label{step41}
				\State $n=1$
				\Repeat 
				\State $\bm{E}(n)=\bm{X}(n)-\bm{\tilde{X}}(n)$\label{algo1:step4} 
				\State $\bm{\xi}^{(n)}=\rho\bm{\xi}^{(n-1)}+\bm{I}_S-\rho\bm{I}_S$
				\State $\bm{Z}^{(n)}=\bm{\Upsilon}^T(n)\bm{P}^{(n-1)}\bm{\Upsilon}^*(n)+\bm{\xi}^{(n)}$\label{step6}
				\State \hspace{-0mm}$\bm{P}^{(n)} \hspace{-1mm}= \hspace{-1mm}(\bm{P}^{(\hspace{-0.5mm}n-1\hspace{-0.5mm})} \hspace{-0.5mm}- \hspace{-0.5mm} \bm{P}^{(\hspace{-0.5mm}n-1\hspace{-0.5mm})}\bm{\Upsilon}^*(\hspace{-0.5mm}n\hspace{-0.5mm}){\bm{Z}^{(\hspace{-0.5mm}n\hspace{-0.5mm})}}^{-1}\bm{\Upsilon}^T\hspace{-0.5mm}(\hspace{-0.5mm}n\hspace{-0.5mm})\bm{P}^{(\hspace{-0.5mm}n-1\hspace{-0.5mm})}){\bm{\Xi}^{(\hspace{-0.5mm}n\hspace{-0.5mm})}}^{-1}$ \label{step7}
				\State $\bm{\tilde{\Phi}}^{(n)} =  \bm{\tilde{\Phi}}^{(n-1)} + \bm{P}^{(n)}\bm{\Upsilon}^*(n)\bm{E}(n)$\label{algo1:step11}
				\State Assign $\bm{\Phi}^{(n)} = \bm{\tilde{\Phi}}^{(n)}$ and compute $\bm{X}(n+1)$ using $\eqref{eq:x_l_FF}$, then find $\bm{\Upsilon}(n+1)$ using the outputs $\bm{Y}(n+1)$ of the PAs\label{step9}
				\State $\bm{\tilde{X}}(n+1) = \bm{\Upsilon}(n+1)^T\bm{\tilde{\Phi}}^{(n)}$\label{algo1:step13}
				\State $n=n+1$
				\Until{$\bm{\tilde{\Phi}}^{(n)}$ converges}
				\State $\bm{\widehat{\Phi}}=\bm{\tilde{\Phi}}^{(n)}$\label{algo1:step10}
			\end{algorithmic}
		}
	\end{algorithm}
	In Algorithm~\ref{algo_1},  the values are assigned to the independent parameters $\rho$, $\lambda_0$, and $\mu$. Then, the initial values of the covariance matrix $\bm{P}$ and the forgetting factor $\bm{\xi}$ are computed in Steps~\ref{step1} and \ref{step2}. Thereafter, the calculations from Step~\ref{algo1:step4} to Step~\ref{algo1:step13} are repeated until the coefficient vector $\bm{\tilde{\Phi}}$ converges. Lastly, we obtain the optimal coefficient vector $\bm{\widehat{\tilde{\Phi}}}$ in Step~\ref{algo1:step10} which is copied to the FF-DPD, i.e., $\bm{\widehat{\Phi}}=\bm{\widehat{\tilde{\Phi}}}$.
	
	\subsubsection*{Performance}
	If we examine Step~\ref{algo1:step11} of Algorithm~\ref{algo_1}, $\bm{\Phi}$ is iteratively estimated using the correlation matrix $\bm{P}$. Therefore, the coefficients in the vector $\bm{\Phi}_l$ of the block vector $\bm{\Phi}$ are correlated with each other to provide optimal predistorted signal $x_l$ to the linearize the $l$th PA. As each PA is provided a separate predistorted signal, hence, FF-DPD gives the best performance. 
	
	\subsubsection*{Complexity} 
	As the matrix multiplication dominates in the complexity of an algorithm~\cite{li16}, therefore, we consider Steps~\ref{step6}, \ref{step7}, and \ref{algo1:step11} to determine the computational complexity of Algorithm~\ref{algo_1} in an iteration. In Step~\ref{step6}, the matrices $\bm{\Upsilon}$, $\bm{P}$ have the sizes $QS\times S$ and $QS\times QS$, respectively. The computational complexity of $\bm{\Upsilon}^T\bm{P}$ is $O(SQSQS)=O(Q^2S^3)$. Now, the matrix $\bm{\Upsilon}^T\bm{P}$ has the size $S\times QS$ which is further multiplied by $\bm{\Upsilon}^*$ with complexity $O(SQSS)=O(QS^3)$. Thus, the total complexity of Step~\ref{step6} is $O(Q^2S^3+QS^3)$. Similarly, the complexities of Steps~\ref{step7} and \ref{algo1:step11} are $O(2Q^3S^3+2Q^2S^3+QS^3)$ and $O(Q^2S^3+QS^2)$, respectively\footnote{Note that the algorithm is still free from the matrix inverse operations, because, although, $\bm{Z}$ and $\bm{\Xi}$ are the matrices, but, they are diagonal matrices and their inverses are only the inverse of their diagonal elements.}. In these operations, the highest order term is $Q^3S^3$, therefore, per iteration, the computational complexity of the algorithm is $O(Q^3S^3)$. 
	
	\begin{lemma}\label{lemma2}
		Using property of diagonal matrix multiplication, complexity of Algorithm~\ref{algo_1} is reduced by a factor of $S^2$.
	\end{lemma}
	\begin{IEEEproof}
		As the direct multiplication of diagonal matrices is not computationally efficient due to redundant multiplication of $0$ entries. For example, the two diagonal matrices of size $2\times 2$ complies the following multiplication, $\text{diag}(a_1,a_2)\text{diag}(b_1,b_2)=\text{diag}(a_1b_1,a_2b_2)$, where $a_i$ and $b_i$; $i\in\{1,2\}$ are the scalars. If we consider the conventional matrix multiplication, total number of multiplications is $2^3$, but, according to the diagonal matrix multiplication, only $2$ multiplications are required of their diagonal entries. Thus, there are $6$ redundant multiplications in the former method. The same multiplication property is applied if $a_i$ and $b_i$ are matrices provided their sizes should satisfy the multiplication  $a_ib_i$. If we employ it in the matrix multiplication $\bm{\Upsilon}^T\bm{P}\bm{\Upsilon}^*$ of Step~\ref{step6}, there are $2S$ respective diagonal elements multiplications. Further, in each diagonal multiplication, for example, in $l$th, the matrix multiplication is $\bm{\Psi}_l^{'T}P_l{\bm{\Psi}_l^{'}}$  with total number of multiplications $O(Q^2+Q)$. Thus, total multiplication is $O(Q^2S+QS)$. Similarly, the computational complexities of Steps~\ref{step7} and \ref{algo1:step11} are $O(2Q^3S+2Q^2S+QS)$ and $O(Q^2S+QS)$, respectively. Thus, per iteration, the computational complexity is $O(Q^3S)$ which is $S^2$ times less than the conventional method.
	\end{IEEEproof}
	
   \subsection{Realization of Learning for LC-DPD using FF-DPD}\label{sec:lern_algo2}
	To learn the LC-DPD coefficient vector $\bm{\overline{\Phi}}$, we exploit the learning process for FF-DPD. In this regard, vector $\bm{\overline{\Phi}}$ is converted into FF-DPD coefficient vector $\bm{\Phi}$ and vice versa. To realize the conversions mathematically, we define two operators $\bm{\mathfrak{M}}_1$ and $\bm{\mathfrak{M}}_2$.
	
	\begin{definition}[\textbf{A Linear Operator $\bm{\mathfrak{M}}_1$}]\label{def2}
		The function $f_1(\cdot)$ that transforms the shape of the coefficient vector $\bm{\overline{\Phi}}$ into the shape of $\bm{\Phi}$ as expressed in~\eqref{eq:f1_op1} is a linear operator $\bm{\mathfrak{M}}_1$ as defined in~\eqref{eq:M_op1}. 
		\begin{subequations}
			\begin{align}\label{eq:f1_op1}
			&\!\!\!\!\!\bm{\Phi} = f_1(\bm{\overline{\Phi}}) = \bm{\mathfrak{M}}_1\bm{\overline{\Phi}},\!\!\\\label{eq:M_op1}
			&\!\!\!\!\! \bm{\mathfrak{M}}_1\hspace{-1mm}\triangleq\hspace{-1mm}\begin{bmatrix}
			\bm{M}_{11}   &\cdots & \bm{M}_{1g}\\
			\vdots  & \ddots & \vdots\\
			\bm{M}_{S1}  & \cdots & \bm{M}_{Sg}
			\end{bmatrix}\!\!,
			\bm{M}_{ij} \hspace{-1mm}= \hspace{-1mm}
			\begin{bmatrix}
			m_{11}^{ij}   &\cdots & m_{1L_j}^{ij}\\
			\vdots  & \ddots & \vdots\\
			m_{Q1}^{ij} & \cdots & m_{QL_j}^{ij}
			\end{bmatrix}\!\!
			\end{align}
		\end{subequations}
		where $L_j=n_jSr^{(\nu+j-1)}$, $m_{uv}^{ij}\in\{0,1\}$ for $u\in\{1,\cdots,Q\}$, $v\in\{1,\cdots,L_j\}$, $i\in\{1,\cdots,S\}$ and $j\in\{1,\cdots,g\}$. Here, $m_{uv}^{ij}=1$ indicates that after performing the operation in~\eqref{eq:f1_op1}, the $v$th element of vector $\bm{\overline{\Phi}}_j$ is assigned to the $u$th element of vector $\bm{\Phi}_i$. Furthermore, the operator $\bm{\mathfrak{M}}_1$ has the following two properties.
		\begin{itemize}
			\item[(i)] In each row vector of the matrix $\bm{\mathfrak{M}}_1$, only one element is $1$ and the remaining elements are $0$.
			\item[(ii)] The sum of the elements in the $z$th column vector of the operator depicts the repetition of $z$th element of vector $\bm{\overline{\Phi}}$ in the vector $\bm{\Phi}$. Also, if the $z$th element of $\bm{\overline{\Phi}}$ lies in the $j$th vector $\bm{\overline{\Phi}}_j$ of the block vector $\bm{\overline{\Phi}}$, then the number of repetition of $z$th coefficient of $\bm{\overline{\Phi}}$ in $\bm{\Phi}$ is $r^{-(\nu+j-1)}$ which is same as the sum of the elements of the $z$th column vector of $\bm{\mathfrak{M}}_1$.
		\end{itemize}
	\end{definition}
	\begin{IEEEexample}
		From Fig.~\ref{fig:sys_mod}(b), the block vector $\bm{\overline{\Phi}}=[\bm{\overline{\Phi}}_1^T,\bm{\overline{\Phi}}_2^T]^T$, where $\bm{\overline{\Phi}}_1 = [\overline{\phi}_{1,1}, \overline{\phi}_{3,1}, \overline{\phi}_{1,2}, \overline{\phi}_{3,2}]^T$, and $\bm{\overline{\Phi}}_2 =[\overline{\phi}_{1,3}, \overline{\phi}_{1,4}]^T$. Here, the parameters for LC-DPD are: $\nu=1$, $r=1/2$, $g=2$, $n_1=n_2=2$, $Q=4$, and $S=4$. To reshape $\bm{\overline{\Phi}}$ into the shape of $\bm{\Phi}=[\bm{\Phi}_1^T,\bm{\Phi}_2^T,\bm{\Phi}_3^T,\bm{\Phi}_4^T]^T$ as for FF-DPD in Fig.~\ref{fig:sys_mod}(a), we use the operator $\mathfrak{M}_1$ as given below, where $\bm{\Phi}_i = [\phi_{i,1},\phi_{i,2},\phi_{i,3},\phi_{i,4}]^T$ for $i\in\{1,2,3,4\}$.
		\begin{align}\label{eq:mat_ex}
		\bm{\mathfrak{M}}_1=\begin{bmatrix}
		\bm{M}_{11} & \bm{M}_{12} \\
		\bm{M}_{21} & \bm{M}_{22}  \\
		\bm{M}_{31} & \bm{M}_{32} \\
		\bm{M}_{41} & \bm{M}_{42} 
		\end{bmatrix}=\footnotesize \left[\begin{array}{@{}c|c@{}}
		\begin{matrix}[0.7]
		1 & 0 & 0 & 0\\
		0 & 0 & 1 & 0\\
		0 & 0 & 0 & 0\\
		0 & 0 & 0 & 0
		\end{matrix}
		& \begin{matrix}[0.7]
		0 & 0 \\
		0 & 0 \\
		1 & 0 \\
		0 & 1
		\end{matrix} \\
		\hline
		\begin{matrix}[0.7]
		1 & 0 & 0 & 0\\
		0 & 0 & 1 & 0\\
		0 & 0 & 0 & 0\\
		0 & 0 & 0 & 0
		\end{matrix}
		& \begin{matrix}[0.7]
		0 & 0 \\
		0 & 0 \\
		1 & 0 \\
		0 & 1
		\end{matrix}\\
		\hline
		\begin{matrix}[0.7]
		0 & 1 & 0 & 0\\
		0 & 0 & 0 & 1\\
		0 & 0 & 0 & 0\\
		0 & 0 & 0 & 0
		\end{matrix}
		& \begin{matrix}[0.7]
		0 & 0 \\
		0 & 0 \\
		1 & 0 \\
		0 & 1
		\end{matrix}\\
		\hline
		\begin{matrix}[0.7]
		0 & 1 & 0 & 0\\
		0 & 0 & 0 & 1\\
		0 & 0 & 0 & 0\\
		0 & 0 & 0 & 0
		\end{matrix}
		& \begin{matrix}[0.7]
		0 & 0 \\
		0 & 0 \\
		1 & 0 \\
		0 & 1
		\end{matrix}
		\end{array}\right]
		\end{align}
		In~\eqref{eq:mat_ex}, $\bm{\mathfrak{M}}_1$ satisfies the property (i) as each of its row vector contains one element as $1$ and remaining are $0$. Next, to verify the property (ii), for instance, the sum of the elements of the $5$th $(z=5)$ column vector is $4$ which entails that $5$th coefficient $\overline{\phi}_{1,3}$ of $\bm{\overline{\Phi}}$ repeats $4$ times in $\bm{\Phi}$. It can also be determined using the expression $r^{-(\nu+j-1)}$ where $j=2$ as the 5th $(z=5)$ column vector lies in the 2nd $(j=2)$ group and it gives $4$ after substitution of the value of the parameters. Besides, the block vector $\bm{\Phi}=[\bm{\Phi}_1^T,\bm{\Phi}_2^T,\bm{\Phi}_3^T,\bm{\Phi}_4^T]^T$ is obtained using the computation in~\eqref{eq:f1_op1}, where $\bm{\Phi}_1=[\overline{\phi}_{1,1},\overline{\phi}_{1,2},\overline{\phi}_{1,3}, \overline{\phi}_{1,4}]^T$, $\bm{\Phi}_2=[\overline{\phi}_{1,1},\overline{\phi}_{1,2},\overline{\phi}_{1,3}, \overline{\phi}_{1,4}]^T$, $\bm{\Phi}_3=[\overline{\phi}_{3,1},\overline{\phi}_{3,2},\overline{\phi}_{1,3}, \overline{\phi}_{1,4}]^T$, and $\bm{\Phi}_4=[\overline{\phi}_{3,1},\overline{\phi}_{3,2},\overline{\phi}_{1,3}, \overline{\phi}_{1,4}]^T$.
	\end{IEEEexample}
	
	\begin{definition}[\textbf{A Linear Operator $\bm{\mathfrak{M}}_2$}]\label{def3}
		The function $f_2(\cdot)$ that transforms the shape of the coefficient vector $\bm{\Phi}$ into the shape of $\bm{\overline{\Phi}}$ as expressed in~\eqref{eq:f2_op2}, where some of the elements of $\bm{\overline{\Phi}}$ are the average of some of elements of $\bm{\Phi}$, is a linear operator $\bm{\mathfrak{M}}_2$ as defined in~\eqref{eq:M2_op2}. 
		\begin{subequations}
			\begin{align}\label{eq:f2_op2}
			&\!\!\!\!\!\bm{\overline{\Phi}} = f_2(\bm{\Phi}) = \bm{\mathfrak{M}}_2\bm{\Phi},\!\!\\\label{eq:M2_op2}
			&\!\!\!\!\! \bm{\mathfrak{M}}_2\hspace{-1mm}\triangleq\hspace{-1mm}\begin{bmatrix}
			\bm{M}_{11}   &\cdots & \bm{M}_{1S}\\
			\vdots  & \ddots & \vdots\\
			\bm{M}_{g1}  & \cdots & \bm{M}_{gS}
			\end{bmatrix}\!\!,
			\bm{M}_{ij} \hspace{-1mm}= \hspace{-1mm}
			\begin{bmatrix}
			m_{11}^{ij}   &\cdots & m_{1Q}^{ij}\\
			\vdots  & \ddots & \vdots\\
			m_{L_i1}^{ij} & \cdots & m_{L_iQ}^{ij}
			\end{bmatrix}\!\!\\\label{eq:mij_val}
			& \!\!\!\!\! \left\{\hspace{-0mm}\begin{array}{cc}
			\hspace{-2mm}m_{uv_1}^{ij_1} =\cdots =  m_{uv_{N_i}}^{ij_{N_i}} = 1/{N_i};&\hspace{-0mm}  \text{for } \overline{\phi}_{i,u}=1/{N_i}\sum_{t=1}^{N_i}\phi_{j_t,v_t}, \\ 
			\hspace{-31.5mm}m_{uv}^{ij}=0; &  \hspace{-28.5mm}\text{ Otherwise}, 
			\end{array}\right.\!\!
			\end{align}
		\end{subequations}
		where $L_i=n_iSr^{(\nu+i-1)}$, $N_i = r^{(\nu+i-1)}$, $u\in\{1,\cdots,L_i\}$, $v,v_1,v_2,\cdots\in\{1,\cdots,Q\}$, $i\in\{1,\cdots,g\}$, and $j\in\{1,\cdots,S\}$. The value of the matrix $\bm{\mathfrak{M}}_2$ elements is determined using~\eqref{eq:mij_val} based on the relationship between the elements of the vectors $\bm{\overline{\Phi}}$ and $\bm{\Phi}$. Besides, the operator $\bm{\mathfrak{M}}_2$ has the following two properties.
		\begin{itemize}
			\item[(i)] In each row vector of the matrix $\bm{\mathfrak{M}}_2$, the sum of the elements is $1$.
			\item[(ii)] Each column vector of $\bm{\mathfrak{M}}_2$ has only one nonzero element which takes the value $1/{N_i}$.
		\end{itemize}
	\end{definition}
	\begin{IEEEexample}
		If we consider Fig.~\ref{fig:sys_mod} for this example, from Fig.~\ref{fig:sys_mod}(a), the block vector $\bm{\Phi}$ for FF-DPD is expressed as $\bm{\Phi}=[\bm{\Phi}_1^T,\bm{\Phi}_2^T,\bm{\Phi}_3^T,\bm{\Phi}_4^T]^T$, where $\bm{\Phi}_j = [\phi_{j,1},\phi_{j,2},\phi_{j,3},\phi_{j,4}]^T$ for $S=4$ and $Q=4$. To transform $\bm{\Phi}$ into the shape of $\bm{\overline{\Phi}}=[\bm{\overline{\Phi}}_1^T,\bm{\overline{\Phi}}_2^T]^T$ as for the LC-DPD in Fig.~\ref{fig:sys_mod}(b), we use the operator $\mathfrak{M}_2$ as given below, where $\bm{\overline{\Phi}}_1 = [\overline{\phi}_{1,1}, \overline{\phi}_{3,1}, \overline{\phi}_{1,2}, \overline{\phi}_{3,2}]^T$, and $\bm{\overline{\Phi}}_2 =[\overline{\phi}_{1,3}, \overline{\phi}_{1,4}]^T$.
		\begin{align}\label{eq:mat_ex2}
		\bm{\mathfrak{M}}_2=\footnotesize 
		\begingroup 
		\setlength\arraycolsep{2pt}
		\left[\begin{array}{@{}c|c|c|c@{}}
		\begin{matrix}[1.3]
		\frac{1}{2} & 0 & 0 & 0\\
		0 & 0 & 0 & 0\\
		0 & \frac{1}{2} & 0 & 0\\
		0 & 0 & 0 & 0
		\end{matrix}
		& \begin{matrix}[1.3]
		\frac{1}{2} & 0 & 0 & 0\\
		0 & 0 & 0 & 0\\
		0 & \frac{1}{2} & 0 & 0\\
		0 & 0 & 0 & 0
		\end{matrix}
		& \begin{matrix}[1.3]
		0 & 0 & 0 & 0\\
		\frac{1}{2} & 0 & 0 & 0\\
		0 & 0 & 0 & 0\\
		0 & \frac{1}{2} & 0 & 0
		\end{matrix}
		& \begin{matrix}[1.3]
		0 & 0 & 0 & 0\\
		\frac{1}{2} & 0 & 0 & 0\\
		0 & 0 & 0 & 0\\
		0 & \frac{1}{2} & 0 & 0
		\end{matrix} \\
		\hline
		\begin{matrix}[1.3]
		0 & 0 & \frac{1}{4} & 0\\
		0 & 0 & 0 & \frac{1}{4}
		\end{matrix}
		& \begin{matrix}[1.3]
		0 & 0 & \frac{1}{4} & 0\\
		0 & 0 & 0 & \frac{1}{4}
		\end{matrix}
		& \begin{matrix}[1.3]
		0 & 0 & \frac{1}{4} & 0\\
		0 & 0 & 0 & \frac{1}{4}
		\end{matrix}
		& \begin{matrix}[1.3]
		0 & 0 & \frac{1}{4} & 0\\
		0 & 0 & 0 & \frac{1}{4}
		\end{matrix}
		\end{array}\right]
		\endgroup
		\end{align}
		In~\eqref{eq:mat_ex2}, $\bm{\mathfrak{M}}_2$ satisfies the property (i) as sum of the elements in each of its row vector is $1$. Further, in each column vector, only one element is nonzero and its value is $1/{N_i}$. For instance, in the second column vector, the third element is nonzero and for it, parameter $i=1$. So, $N_1=(1/2)^{(1+1-1)}=1/2$. Thus, property (ii) is also satisfied by $\bm{\mathfrak{M}}_2$. Moreover, after performing the operation in~\eqref{eq:M2_op2}, the relationship between the elements of $\bm{\overline{\Phi}}$ and $\bm{\Phi}$ can be expressed as: $\overline{\phi}_{1,1} = (\phi_{1,1}+\phi_{2,1})/2$, $\overline{\phi}_{3,1} = (\phi_{3,1}+\phi_{4,1})/2$, $\overline{\phi}_{1,2} = (\phi_{1,2}+\phi_{2,2})/2$, $\overline{\phi}_{3,2} = (\phi_{3,2}+\phi_{4,2})/2$, $\overline{\phi}_{1,3} = (\phi_{1,3}+\phi_{2,3}+\phi_{3,3}+\phi_{4,3})/4$, and $\overline{\phi}_{1,4} = (\phi_{1,4}+\phi_{2,4}+\phi_{3,4}+\phi_{4,4})/4$.
	\end{IEEEexample} 
	
	\begin{algorithm}[!t]
		{\small
			\caption{\small Estimation of coefficients for LC-DPD (Method-I).}\label{algo_2}
			\begin{algorithmic}[1]
				\Require The values of $\rho$, $\lambda_0$, $\mu$, $\bm{\mathfrak{M}}_1$, $\bm{\mathfrak{M}}_2$, and $\bm{\tilde{\overline{\Phi}}}^{(0)}$
				\Ensure The estimated coefficient vector $\bm{\widehat{\overline{\Phi}}}$
				\State Determine the initial values $\bm{P}^{(0)}$ and $\bm{\xi}^{(0)}$ using Steps~\ref{step1} and~\ref{step2} of Algorithms~\ref{algo_1}
				\State Find $\bm{\tilde{\Phi}}^{(0)}=\mathfrak{M}_1\bm{\tilde{\overline{\Phi}}}^{(0)}$, then, assign $\bm{\Phi}^{(0)}=\bm{\tilde{\Phi}}^{(0)}$ and compute $\bm{X}(1)$, $\bm{\Upsilon}(1)$, and $\bm{\tilde{X}}(1)$ as in Steps~\ref{step3} and~\ref{step41} of Algorithm~\ref{algo_1}\label{algo2:step2}
				\State $n=1$
				\Repeat 
				\State Compute $\bm{E}(n)$, $\bm{\xi}^{(n)}$, $\bm{Z}^{(n)}$, $\bm{P}^{(n)}$, and $\bm{\tilde{\Phi}}^{(n)}$ from Step~\ref{algo1:step4} to Step~\ref{algo1:step11} of Algorithm~\ref{algo_1}.
				\State Using operator $\bm{\mathfrak{M}}_2$ in~\eqref{eq:M2_op2}, compute $\bm{\tilde{\overline{\Phi}}}^{(n)} = \bm{\mathfrak{M}}_2 \bm{\tilde{\Phi}}^{(n)}$\label{algo2:step6}
				\State Then, using $\bm{\mathfrak{M}}_1$ in~\eqref{eq:M_op1}, compute $\bm{\tilde{\Phi}}^{(n)} = \bm{\mathfrak{M}}_1 \bm{\tilde{\overline{\Phi}}}^{(n)}$\label{algo2:step7}
				\State Assign $\bm{\Phi}^{(n)} = \bm{\tilde{\Phi}}^{(n)}$ and compute $\bm{X}(n+1)$,  $\bm{\Upsilon}(n+1)$ as in Step~\ref{step9} of Algorithm~\ref{algo_1}
				\State $\bm{\tilde{X}}(n+1) = \bm{\Upsilon}(n+1)^T\bm{\tilde{\Phi}}^{(n)}$
				\State $n=n+1$
				\Until{$\bm{\tilde{\overline{\Phi}}}^{(n)}$ converges}
				\State $\bm{\widehat{\overline{\Phi}}}=\bm{\tilde{\overline{\Phi}}}^{(n)}$
			\end{algorithmic}
		}
	\end{algorithm}
   Now, using Algorithm~\ref{algo_2}, we train the coefficient vector $\bm{\tilde{\overline{\Phi}}}$ of LC-DPD as follows. Apart from the parameters, $\rho$, $\lambda_0$, $\mu$, and $\bm{\tilde{\overline{\Phi}}}^{(0)}$, the values of the operators, $\bm{\mathfrak{M}}_1$ and $\bm{\mathfrak{M}}_2$ are also inputted to the algorithm. As we realize the training of $\bm{\tilde{\overline{\Phi}}}$ by exploiting the training of $\bm{\tilde{\Phi}}$, the steps of Algorithm~\ref{algo_2} are same as of Algorithm~\ref{algo_1} except the Steps \ref{algo2:step2}, \ref{algo2:step6}, and \ref{algo2:step7}. In Step~\ref{algo2:step2}, using operator $\mathfrak{M}_1$, $\bm{\tilde{\overline{\Phi}}}^{(0)}$ is converted into $\bm{\tilde{\Phi}}^{(0)}$ to compute other initial values of the parameters for the FF-DPD structure. While Steps \ref{algo2:step6} and  \ref{algo2:step7} enforce the learning of FF-DPD coefficients in $\bm{\tilde{\Phi}}$ to incorporate the repetitive characteristics of LC-DPD coefficients in  $\bm{\tilde{\overline{\Phi}}}$ in each iteration. The forth process in Step~\ref{algo2:step6} takes the average of some coefficients of $\bm{\tilde{\Phi}}$ which has to be repeated in the LC-DPD structure and is assigned to a coefficient of $\bm{\tilde{\overline{\Phi}}}$ (cf. example of Definition~\ref{def3}). In the back process  in Step~\ref{algo2:step7}, again, this coefficient in $\bm{\tilde{\overline{\Phi}}}$ is repeated in $\bm{\tilde{\Phi}}$ (cf. example of Definition~\ref{def2}). Thus, it enforces the repetitive coefficients in $\bm{\tilde{\Phi}}$ to have equal value in each iteration as the learning is based on FF-DPD. Finally, after the convergence, it returns the estimated LC-DPD coefficient vector $\bm{\widehat{\overline{\Phi}}}$. 
   
   \subsubsection*{Performance and Complexity}
   As in Algorithm~\ref{algo_2}, the LC-DPD coefficient vector $\bm{\overline{\Phi}}$ is trained by exploiting the training of the FF-DPD coefficient vector $\bm{\Phi}$ where using the operator $\mathfrak{M}_2$, the common coefficients in $\bm{\overline{\Phi}}$ is obtained by taking the average of some of the coefficients in $\bm{\Phi}$ (cf. Example of Definition~\ref{def3}). However, the obtained common coefficients after the average loose the correlation with the distinct coefficients. Therefore, using it, the generated predistorted signals are not optimal as in FF-DPD to linearize the PAs; thus, its performance is low. Further, the complexity of the algorithm is described as follows. Although, the operators, $\mathfrak{M}_1$ and $\mathfrak{M}_2$ are represented as the matrices in \eqref{eq:M_op1} and \eqref{eq:M2_op2} to analyze the operations mathematically, but, in practice, they have only assignment and average operations whose complexities are negligible compared to the matrix multiplications. Therefore, the dominant operations in Algorithm~\ref{algo_2} are same as Algorithm~\ref{algo_1}, thus, its complexity per iteration is $O(Q^3S)$. To enhance the performance and to reduce the complexity, we propose the improved algorithms in next section. 
  
  \section{Training Based on ILA-RPEM: Part II}	
  For reducing the complexity of the algorithm, we need to train LC-DPD coefficient vector $\bm{\overline{\Phi}}$ by only exploiting the structure of LC-DPD instead of enforcing its training using the FF-DPD. Because, the length of vector $\bm{\overline{\Phi}}$ as given by~\eqref{eq:N_m} is less than $\bm{\Phi}$. Thus, the training only using the length of vector $\bm{\overline{\Phi}}$ reduces the sizes of the matrices, $\bm{P}$ and $\bm{\Upsilon}$ in the dominant matrix multiplications. Based on it, we propose two algorithms.
  
  In order to train $\bm{\overline{\Phi}}$ by completely exploiting the LC-DPD structure, first, we need to represent $\bm{\overline{\Phi}}$ in a suitable form, i.e., in another block vector where each vector in it consists the coefficients that are multiplied by the BFs to generate a predistorted signal that is distributed to a subgroup of PAs\footnote{In LC-DPD assisted subarray, the $S$ PAs of the subarray are divided into $\overline{\sigma}_g=Sr^\nu$ subgroups, where each subgroup consists $n_{PA}=S/\overline{\sigma}_g=r^{-\nu}$ number of PAs. Thus, LC-DPD structure generates $\overline{\sigma}_g$ predistorted signals to distribute them to respective subgroups.}.
  
  \begin{definition}[\textbf{Reshape of $\bm{\overline{\Phi}}$ as $\bm{\overline{\Phi}^{'}}$}]
  	   To generate the predistorted signals using the LC-DPD coefficient vector $\bm{\overline{\Phi}}$, it can be reshaped as the block vector $\bm{\overline{\Phi}^{'}}$, given by:
  	   \begin{align}\label{eq:def4}
  	   \bm{\overline{\Phi}}^{'} = [\underbrace{\bm{\overline{\Phi}}_{1}^{'T}, \cdots, \bm{\overline{\Phi}}_{\overline{\sigma}_1}^{'T}}_{1st \text{ gr., each of len. }J_1},\cdots,\underbrace{\bm{\overline{\Phi}}_{(\overline{\sigma}_{g-1}+1)}^{'T}, \cdots, \bm{\overline{\Phi}}_{\overline{\sigma}_g}^{'T}}_{gth \text{ gr., each of len. }J_g}]^T,
  	   \end{align}
  	   where $\bm{\overline{\Phi}}_{i}^{'}$ is the $i$th vector in $\bm{\overline{\Phi}}^{'}$. Here, the grouping of the vectors is based on their lengths, i.e., the vectors in a group have same length.  The total number of groups is $g$ which is equal to the number of vectors in $\bm{\overline{\Phi}}$ (cf. \eqref{eq:block_vec}). In the $j$th group, the number of vectors is $T_j$ and each of the vector is of $J_j$ length as shown in \eqref{eq:def4}. Besides, $\overline{\sigma}_j=\sum_{q=1}^{j}T_q$. Further, $T_j$ and $J_j$ are given by:
  	   \begin{subequations}\label{eq:lemma3_ab}
  	   	\begin{align}
  	   	T_j &= Sr^{\nu+g-j}\left[1-r\tilde{u}\left(r^{1-j}-1\right)\right],\\
  	   	J_j&= Q - \sum_{q=1}^{j-1}n_{g-q+1}\tilde{u}\left(r^{1-j}-1\right),
  	   	\end{align}
  	   \end{subequations}
  	   where $j\in{\{1,\cdots,g\}}$, $r<1$ and $\tilde{u}(x)=1$ for $x>0$; otherwise, $\tilde{u}(x)=0$.
  \end{definition}
  \begin{IEEEexample}
  	Again, we consider the instance of LC-DPD structure with $S=4$, $Q=4$, $r=1/2$, and $\nu =1$ in Fig.~\ref{fig:sys_mod}(b) to reshape $\bm{\overline{\Phi}}$ to $\bm{\overline{\Phi}^{'}}$. From \eqref{eq:block_vec}, $\bm{\overline{\Phi}}=[\bm{\overline{\Phi}}_1^T,\bm{\overline{\Phi}}_2^T]^T$ where $\bm{\overline{\Phi}}_1 = [\underbrace{\overline{\phi}_{1,1}, \overline{\phi}_{3,1}, \overline{\phi}_{1,2}, \overline{\phi}_{3,2}}_{n_1=4}]^T$, $\bm{\overline{\Phi}}_2 =[\underbrace{\overline{\phi}_{1,3}, \overline{\phi}_{1,4}}_{n_2=2}]^T$, and $g=2$. Using \eqref{eq:def4}, its reshape as $\bm{\overline{\Phi}^{'}}$ to generate the predistored signals is given as: $\bm{\overline{\Phi}^{'}}=[\underbrace{\bm{\overline{\Phi}}_{1}^{'T}}_{1\text{st} \text{ gr.}},\underbrace{\bm{\overline{\Phi}}_{2}^{'T}}_{2\text{nd} \text{ gr.}}]^T$. Here, number of groups, $g=2$ and substituting the parameters values in \eqref{eq:lemma3_ab}, we get $T_1=T_2=1$, $J_1=4$, and $J_2=2$. Thus, $\bm{\overline{\Phi}}_{1}^{'}=[\overline{\phi}_{1,1},\overline{\phi}_{1,2},\overline{\phi}_{1,3}, \overline{\phi}_{1,4}]^T$ and $\bm{\overline{\Phi}}_{2}^{'}=[\overline{\phi}_{3,1},\overline{\phi}_{3,2}]^T$.
  \end{IEEEexample}
  	 \begin{corollary}\label{corollary1}
  		As $\bm{\overline{\Phi}^{'}}$ is the reshape of $\bm{\overline{\Phi}}$, therefore, the elements in the former vector is same as in the later, thus, the length of the two vectors is equal. The length of $\bm{\overline{\Phi}^{'}}$ can be obtained by $\sum_{i=1}^{g}T_i J_i$, hence, from the length of $\bm{\overline{\Phi}}$ in \eqref{eq:N_m_a}, $N_m^L=\sum_{i=1}^{g}T_i J_i$.
  	\end{corollary}
  	\noindent The above corollary can be proved by substituting $Q=\sum_{i=1}^{g}n_i$ in \eqref{eq:lemma3_ab} followed by simplifying $\sum_{i=1}^{g}T_i J_i$, it gives $N_m^L$ in \eqref{eq:N_m_a}. Furthermore, to reshape $\bm{\overline{\Phi}}$ into $\bm{\overline{\Phi}^{'}}$, we define a linear operator $\bm{\mathfrak{M}}_3$ as below.
  
  \begin{definition}[\textbf{A Linear Operator $\bm{\mathfrak{M}}_3$}]\label{eq:def5}
  	A linear operator  $\bm{\mathfrak{M}}_3$ is defined as in \eqref{eq:M_3_def} which is used to reshape the LC-DPD coefficient vector $\bm{\overline{\Phi}}$ into the vector $\bm{\overline{\Phi}^{'}}$ as given by \eqref{eq:def4}.  
  	\begin{subequations}\label{eq:M_3}
  		\begin{align}\label{eq:M_3_conv}
  		&\bm{\overline{\Phi}^{'}} = \bm{\mathfrak{M}}_3\bm{\overline{\Phi}},\\\label{eq:M_3_def}
  		&\bm{\mathfrak{M}}_{3} = 
  		\begin{bmatrix}
  		\bm{M}_{11}   &\cdots & \bm{M}_{1g}\\
  		\vdots  & \ddots & \vdots\\
  		\bm{M}_{\overline{\sigma}_11}  & \cdots & \bm{M}_{\overline{\sigma}_1g}\\
  		\vdots  & \ddots & \vdots\\
  		\bm{M}_{(\overline{\sigma}_{g-1}+1)1}   &\cdots & \bm{M}_{(\overline{\sigma}_{g-1}+1)g}\\
  		\vdots  & \ddots & \vdots\\
  		\bm{M}_{\overline{\sigma}_{g}1}  & \cdots & \bm{M}_{\overline{\sigma}_{g}g}
  		\end{bmatrix},\\
  		& \bm{M}_{ij} = 
  		\begin{bmatrix}
  		m_{11}^{ij}   &\cdots & m_{1L_j}^{ij}\\
  		\vdots  & \ddots & \vdots\\
  		m_{J_\tau1}^{ij} & \cdots & m_{J_\tau L_j}^{ij}
  		\end{bmatrix},
  		\end{align}
  	\end{subequations}
  	where $T_\tau$ and $J_\tau$ is given by \eqref{eq:lemma3_ab} for $\tau\in\{1,\cdots,g\}$ , $L_j=n_jSr^{(\nu+j-1)}$, $m_{uv}^{ij}\in\{0,1\}$ for $i\in\{\overline{\sigma}_{\tau-1}+1,\cdots,\overline{\sigma}_{\tau}\}$ $(\overline{\sigma}_0=1)$, $j\in\{1,\cdots,g\}$, $u\in\{1,\cdots,J_\tau\}$, and $v\in\{1,\cdots,L_j\}$. Moreover, sum of the elements in each of the row or column vector of $\bm{\mathfrak{M}}_3$ is $1$.
  \end{definition} 
  \begin{IEEEexample}
  	Again, we consider the example for Fig.~\ref{fig:sys_mod}(b) to reshape $\bm{\overline{\Phi}}=[\bm{\overline{\Phi}}_1^T,\bm{\overline{\Phi}}_2^T]^T$ into $\bm{\overline{\Phi}^{'}}=[\bm{\overline{\Phi}}_1^{'T},\bm{\overline{\Phi}}_2^{'T}]^T$, where $\bm{\overline{\Phi}}_1 = [\overline{\phi}_{1,1}, \overline{\phi}_{3,1}, \overline{\phi}_{1,2}, \overline{\phi}_{3,2}]^T$, $\bm{\overline{\Phi}}_2 =[\overline{\phi}_{1,3}, \overline{\phi}_{1,4}]^T$, $\bm{\overline{\Phi}}_{1}^{'}=[\overline{\phi}_{1,1},\overline{\phi}_{1,2},\overline{\phi}_{1,3}, \overline{\phi}_{1,4}]^T$, and $\bm{\overline{\Phi}}_{2}^{'}=[\overline{\phi}_{3,1},\overline{\phi}_{3,2}]^T$. For it, the operator $\bm{\mathfrak{M}}_3$ is given by:
  	\begin{align}\label{eq:mat_ex_m3}
  	\bm{\mathfrak{M}}_3=\begin{bmatrix}
  	\bm{M}_{11} & \bm{M}_{12} \\
  	\bm{M}_{21} & \bm{M}_{22}
  	\end{bmatrix}=\footnotesize \left[\begin{array}{@{}c|c@{}}
  	\begin{matrix}[0.7]
  	1 & 0 & 0 & 0\\
  	0 & 0 & 1 & 0\\
  	0 & 0 & 0 & 0\\
  	0 & 0 & 0 & 0
  	\end{matrix}
  	& \begin{matrix}[0.7]
  	0 & 0 \\
  	0 & 0 \\
  	1 & 0 \\
  	0 & 1
  	\end{matrix} \\
  	\hline
  	\begin{matrix}[0.7]
  	0 & 1 & 0 & 0\\
  	0 & 0 & 0 & 1
  	\end{matrix}
  	& \begin{matrix}[0.7]
  	0 & 0 \\
  	0 & 0 
  	\end{matrix}
  	\end{array}\right]
  	\end{align}
  \end{IEEEexample}
  \begin{corollary}\label{corollary2}
  	The operator $\bm{\mathfrak{M}}_3$ is always a square matrix as it is used to reshape the vector $\bm{\overline{\Phi}}$ into the vector $\bm{\overline{\Phi}^{'}}$ using the same elements. Also, the column vectors in $\bm{\mathfrak{M}}_3$ are unit vectors and they are orthogonal to each other. Therefore, they form a orthonormal basis in the space $\mathbb{R}^{N_m^L}$. Furthermore, the inverse of $\bm{\mathfrak{M}}_3$ is its transpose, i.e., $\bm{\mathfrak{M}}_3^{-1}=\bm{\mathfrak{M}}_3^T$~\cite{str48}. Hence, from \eqref{eq:M_3_conv}, using $\bm{\mathfrak{M}}_3^{-1}$,  $\bm{\overline{\Phi}^{'}}$ can be reshaped back to $\bm{\overline{\Phi}}$ as: $\bm{\overline{\Phi}}=\bm{\mathfrak{M}}_3^T\bm{\overline{\Phi}^{'}}$.
  \end{corollary}

	\begin{algorithm}[!t]
		{\small
			\caption{\small Estimation of coefficients for LC-DPD (Method-II).}\label{algo_3}
			\begin{algorithmic}[1]
				\Require The values of $\rho$, $\lambda_0$, $\mu$, $\bm{\mathfrak{M}}_1$, $\bm{\mathfrak{M}}_2$, $\bm{\mathfrak{M}}_3$, and $\bm{\tilde{\overline{\Phi}}}^{(0)}$
				\Ensure The estimated coefficients $\bm{\widehat{\overline{\Phi}}}$
				\State $\bm{\overline{P}}^{(0)}=\text{diag}(\underbrace{\mu\bm{I}_{J_1},\cdots,\mu\bm{I}_{J_1}}_{T_1 \text{ times}},\cdots,\underbrace{\mu\bm{I}_{J_g},\cdots,\mu\bm{I}_{J_g}}_{T_g \text{ times}})$\label{M2:step1}
				\State $\bm{\overline{\xi}}^{(0)}=\lambda_0\bm{I}_{\overline{\sigma}_{g}}$\label{M2:step2}
				\State Operate $\bm{\tilde{\Phi}}^{(0)} = \bm{\mathfrak{M}}_1\bm{\tilde{\overline{\Phi}}}^{(0)}$  and  assign $\bm{\Phi}^{(0)}=\bm{\tilde{\Phi}}^{(0)}$\label{algo3:step3} 
				\State Obtain $\bm{X}(1)$ using~\eqref{eq:x_l_FF} followed by $\bm{Y}(1)$ using~\eqref{eq:out_pa}, then, compute $\bm{\Psi}^{'}$, $\bm{\Upsilon}(1)$ similar to Step~\ref{step3} of Algorithm~\ref{algo_1}
				\State Compute $\bm{{\overline{\Psi}^{'}}}(1) = \bm{\mathfrak{M}}_3(\bm{\mathfrak{M}}_2 \bm{\Psi}^{'}(1))$ and $\bm{\overline{\Upsilon}}(1)$\label{algo3:step4}
				\State Compute $\bm{\tilde{X}}(1) = \bm{\Upsilon}(1)^T\bm{\tilde{\Phi}}^{(0)}$ and $\bm{{\tilde{\overline{\Phi}}^{'}}}^{(0)} = \mathfrak{M}_3\bm{\tilde{\overline{\Phi}}}^{(0)}$\label{M2:step41}
				\State $n=1$
				\Repeat 
				\State $\bm{E}(n)=\bm{X}(n)-\bm{\tilde{X}}(n)$\label{M2:step4} 
				\State $\bm{\overline{E}}(n) = \bm{\mathfrak{M}}_3(\bm{\mathfrak{M}}_2(\bm{E}(n)) \bigotimes \bm{1}_Q)$
				\State $\bm{\overline{\xi}}^{(n)}=\rho\bm{\overline{\xi}}^{(n-1)}+\bm{I}_{\overline{\sigma}_{g}}-\rho\bm{I}_{\overline{\sigma}_{g}}$
				\State $\bm{\overline{Z}}^{(n)}=\bm{\overline{\Upsilon}}^T(n)\bm{\overline{P}}^{(n-1)}\bm{\overline{\Upsilon}}^*(n)+\bm{\overline{\xi}}^{(n)}$\label{M2:step6}
				\State \hspace{-0mm}$\bm{\overline{P}}^{(n)} \hspace{-1mm}= \hspace{-1mm}(\bm{\overline{P}}^{(\hspace{-0.5mm}n-1\hspace{-0.5mm})} \hspace{-0.5mm}- \hspace{-0.5mm} \bm{\overline{P}}^{(\hspace{-0.5mm}n-1\hspace{-0.5mm})}\bm{\overline{\Upsilon}}^*(\hspace{-0.5mm}n\hspace{-0.5mm}){\bm{\overline{Z}}^{(\hspace{-0.5mm}n\hspace{-0.5mm})}}^{-1}\bm{\overline{\Upsilon}}^T\hspace{-0.5mm}(\hspace{-0.5mm}n\hspace{-0.5mm})\bm{\overline{P}}^{(\hspace{-0.5mm}n-1\hspace{-0.5mm})}){\bm{\overline{\Xi}}^{(\hspace{-0.5mm}n\hspace{-0.5mm})}}^{-1}$ \label{M2:step7}
				\State $\bm{{\tilde{\overline{\Phi}}^{'}}}^{(n)} =  \bm{{\tilde{\overline{\Phi}}^{'}}}^{(n-1)} + (\bm{\overline{P}}^{(n)}\bm{\overline{\Upsilon}}^*(n)\bm{1}_{\overline{\sigma}_{g}})\bigodot\bm{\overline{E}}(n)$\label{M2:step8}
				\State Operate $\bm{\tilde{\overline{\Phi}}}^{(n)} = \mathfrak{M}_3^T\bm{{\tilde{\overline{\Phi}}^{'}}}^{(n)}$ followed by $\bm{\tilde{\Phi}}^{(n)} = \bm{\mathfrak{M}}_1\bm{\tilde{\overline{\Phi}}}^{(n)}$, then,  assign $\bm{\Phi}^{(n)}=\bm{\tilde{\Phi}}^{(n)}$ 
				\State Using obtained $\bm{X}(n+1)$ followed by $\bm{Y}(n+1)$ and then $\bm{\Psi}^{'}(n+1)$, find $\bm{\Upsilon}(n+1)$, $\bm{{\overline{\Psi}^{'}}}(n+1) =\bm{\mathfrak{M}}_3(\bm{\mathfrak{M}}_2 \bm{\Psi}^{'}(n+1))$, and $\bm{\overline{\Upsilon}}(n+1)$\label{algo3:step15}
				\State $\bm{\tilde{X}}(n+1) = \bm{\Upsilon}(n+1)^T\bm{\tilde{\Phi}}^{(n)}$
				\State $n=n+1$
				\Until{$\bm{\tilde{\overline{\Phi}}}^{(n)}$ converges}
				\State $\bm{\widehat{\overline{\Phi}}}=\bm{\tilde{\overline{\Phi}}}^{(n)}$
			\end{algorithmic}
		}
	\end{algorithm}
	Now, we utilize the operator $\bm{\mathfrak{M}}_3$ to train the coefficient vector $\bm{\overline{\Phi}}$ in Algorithm~\ref{algo_3}. In the algorithm based on ILA-RPEM, the sizes of different matrices and vectors used in the training are determined according to the shape of the vector $\bm{\overline{\Phi}^{'}}$ in \eqref{eq:def4}. They are defined as: the forgetting vector $\bm{\overline{\xi}}\triangleq\text{diag}(\xi_1,\cdots,\xi_{\overline{\sigma}_{g}})$, $\bm{\overline{Z}}\triangleq \text{diag}(Z_1,\cdots,Z_{\overline{\sigma}_{g}})$, $\bm{\overline{P}}\triangleq \text{diag}(P_1,\cdots,P_{\overline{\sigma}_{g}})$, and $\bm{\Xi} \triangleq \text{diag}(\xi_1\bm{I}_{J_1},\cdots,\xi_{\overline{\sigma}_1}\bm{I}_{J_1},\cdots,\xi_{(\overline{\sigma}_{g-1}+1)}\bm{I}_{J_g},\cdots,\xi_{\overline{\sigma}_{g}}\bm{I}_{J_g})$. To determine the parameter $\bm{\overline{\Upsilon}}$, first, using the outputs $\bm{Y}$ of the PAs, we obtain the block vector $\bm{\Psi}^{'}=[\bm{\Psi}_1^{'T},\cdots,\bm{\Psi}_S^{'T}]^T$, where $\bm{\Psi}_l^{'}=[\psi_{l,1}^{'},\cdots,\psi_{l,Q}^{'}]^T$.	 From $\bm{\Psi}^{'}$, the block vector $\bm{\overline{\Psi^{'}}}$ for LC-DPD can be obtained using the operator $\bm{\mathfrak{M}}_2$ as: $\bm{\overline{\Psi^{'}}} = \bm{\mathfrak{M}}_2 \bm{\Psi}^{'}$. Here, $\bm{\overline{\Psi^{'}}} = [\bm{\overline{\Psi^{'}}}_1^T,\cdots,\bm{\overline{\Psi^{'}}}_g^T]^T$ and $\bm{\overline{\Psi^{'}}}_i$ can be represented similar to $\bm{\overline{\Phi}}_i$ in~\eqref{eq:Phi_j} for $i\in\{1,\cdots,g\}$, where $\phi$ is replaced by $\psi^{'}$. Further, using operator $\bm{\mathfrak{M}}_3$, $\bm{\overline{\Psi^{'}}}$ can be reshaped into $\bm{\overline{\Psi}^{'}}$ as:  $\bm{\overline{\Psi}^{'}}=\bm{\mathfrak{M}}_3\bm{\overline{\Psi^{'}}}$. Now, using the vectors in the block vector $\bm{\overline{\Psi}^{'}}$, $\bm{\overline{\Upsilon}}$ can be expressed as\footnote{This procedure is used to compute $\bm{\overline{\Upsilon}}$ from $\bm{Y}$ in Steps~\ref{algo3:step4} and \ref{algo3:step15} of Algorithm~\ref{algo_3}.}: $\bm{\overline{\Upsilon}} = \text{diag}(\bm{\overline{\Psi}^{'}}_1,\cdots,\bm{\overline{\Psi}^{'}}_{\overline{\sigma}_g})$. Now the process in the Algorithm~\ref{algo_3} can be described as follows. The inputs to the algorithm are same as in Algorithm~\ref{algo_2} along with the operator $\bm{\mathfrak{M}}_3$. In the first two steps, correlation matrix $\bm{\overline{P}}$ and $\bm{\overline{\xi}}$ are initialized. Using Steps \ref{algo3:step3} to \ref{M2:step41}, the initial values of $\bm{\overline{\Upsilon}}$ and the postdistorted signal vector $\bm{\tilde{X}}$ are determined. Thereafter, similar to Algorithms \ref{algo_1} and \ref{algo_2}, the iterative steps are followed to get the the converged value of $\bm{\widehat{\overline{\Phi}}}$. In the loop, the operators, $\bigotimes$ and $\bigodot$ represent the Kronecker and Hadamard products, respectively. 
	
	\subsubsection*{Performance and Complexity}
	In Algorithm~\ref{algo_3}, the block vector $\bm{\overline{\Phi}}$ is reshaped as the block vector $\bm{{\overline{\Phi}}^{'}}$ to correlate the coefficients in its vector $\bm{{\overline{\Phi}}^{'}}_i$ using the correlation matrix $\bm{\overline{P}}$ while the training (cf. Step~\ref{M2:step8} of Algorithm~\ref{algo_3}). However, the common coefficients in it are correlated with the distinct coefficients in the first $\overline{\sigma}_1=T_1$ vectors of the block vector $\bm{{\overline{\Phi}}^{'}}$ (cf. \eqref{eq:def4}). For example, in Fig.~\ref{fig:sys_mod}(b), the coefficients, $\overline{\phi}_{1,3}$, $\overline{\phi}_{1,4}$ are commonly shared with the remaining coefficients $\overline{\phi}_{1,1}$, $\overline{\phi}_{1,2}$ and $\overline{\phi}_{3,1}$, $\overline{\phi}_{3,2}$ to generate the predistorted signals $x_1$ and $x_2$, respectively. But, using Algorithm~\ref{algo_3}, $\overline{\phi}_{1,3}$, $\overline{\phi}_{1,4}$ are only correlated with $\overline{\phi}_{1,1}$, $\overline{\phi}_{1,2}$, thus, the predistorted signal $x_1$ is better to linearize $1$st subgroup of PAs, whereas, $x_2$ performs less to linearize the $2$nd subgroup. Therefore, the predistorted signals generated using the coefficients in first $T_1$ vectors in $\bm{{\overline{\Phi}}^{'}}$ are optimal to linearize the respective subgroups of PAs. But, still, this algorithm provides a low performance for the remaining PAs. Therefore, next, we propose an algorithm that establishes the correlation of the common coefficients with the all distinct coefficients. Besides, the complexity of Algorithm~\ref{algo_3} in an  iteration can be determined using the dominant matrix multiplications in Steps \ref{M2:step6}, \ref{M2:step7}, and \ref{M2:step8}. As the matrices, $\bm{\overline{\Upsilon}}$, $\bm{\overline{P}}$, $\bm{\overline{Z}}$, and $\bm{\overline{\Xi}}$ are diagonal, the respective complexities of the three steps are: $O(Q^2T_1+QT_1)$, $O(2Q^3T_1+2Q^2T_1+QT_1)$, and $O(Q^2T_1+QT_1)$. Based on the dominant term, the complexity of Algorithm~\ref{algo_3} is $O(Q^3T_1)$. As $T_1<S$, the complexity of Algorithm~\ref{algo_3} is lesser than Algorithms \ref{algo_1} and \ref{algo_2}. Next, to correlate the common coefficients in LC-DPD structure with all the remaining, first, we define a sequence of linear operators as below.
	
	\begin{definition}[\textbf{A Sequence of Linear Operators for the Back and Forth Operations}]
		To establish the correlation of the common coefficients to $r^{-(g-1)}$ set of distinct coefficients in the LC-DPD structure, a sequence of $r^{-(g-1)}$ operators is defined in \eqref{eq:seq_op}, where the $t$th operator $\bm{\mathfrak{M}}_{4,t}$ in the sequence is given by \eqref{eq:seq_op_t} and its $ij$th matrix element $ \bm{M}_{ij,t}$ is expressed in \eqref{eq:seq_ele_ijt}.
		\begin{subequations}
			\begin{align}\label{eq:seq_op}
				& \bm{\mathfrak{M}}_{4} = \{\bm{\mathfrak{M}}_{4,1},\bm{\mathfrak{M}}_{4,2},\cdots,\bm{\mathfrak{M}}_{4,r^{-(g-1)}}\}\\\label{eq:seq_op_t}
				&\bm{\mathfrak{M}}_{4,t} = 
				\begin{bmatrix}
				\bm{M}_{11,t}   &\cdots & \bm{M}_{1g,t}\\
				\vdots  & \ddots & \vdots\\
				\bm{M}_{T_11,t}  & \cdots & \bm{M}_{T_1g,t}\\
				\bm{M}_{(T_1+1)1,t}  & \cdots & \bm{M}_{(T_1+1)g,t}
				\end{bmatrix}\\\label{eq:seq_ele_ijt}
				& \bm{M}_{ij,t} = 
				\begin{bmatrix}
				m_{11,t}^{ij}   &\cdots & m_{1L_j,t}^{ij}\\
				\vdots  & \ddots & \vdots\\
				m_{\overline{L}_i1,t}^{ij} & \cdots & m_{\overline{L}_iL_j,t}^{ij}
				\end{bmatrix},\\\label{eq:20d}
				& \bm{\Phi}_{tr.,t}^{(n+1)} = \text{trunc}(\bm{\mathfrak{M}}_{4,t}\bm{\overline{\Phi}}^{(n)},Q),\\\label{eq:20e}
				& \bm{\overline{\Phi}}^{(n+1)} = \bm{\mathfrak{M}}_{4,t}^{T}\text{merge}(\bm{\Phi}_{tr.,t}^{(n+1)},\bm{\mathfrak{M}}_{4,t}\bm{\overline{\Phi}}^{(n)},Q)
			\end{align}
		\end{subequations}
		where $t\in\{1,\cdots,r^{-(g-1)}\}$, $\overline{L}_i=Q$ for $i\in{\{1,\cdots,T_1\}}$; otherwise $\overline{L}_i=N_m^L-T_1Q$ for $i=T_1+1$. $L_j=n_jSr^{(\nu+j-1)}$, $m_{uv,t}^{ij}\in\{0,1\}$ for $i\in\{1,\cdots,T_1+1\}$, $j\in\{1,\cdots,g\}$, $u\in\{1,\cdots,\overline{L}_i\}$, and $v\in\{1,\cdots,L_j\}$. Again, like $\bm{\mathfrak{M}}_3$ in \eqref{eq:M_3_def}, the sum of the elements in each of the row or column vector of $\bm{\mathfrak{M}}_{4,t}$ is $1$. Further, from Corollary~\ref{corollary2}, $\bm{\mathfrak{M}}_{4,t}^{-1}=\bm{\mathfrak{M}}_{4,t}^{T}$. Besides, the common coefficients are with the $t$th set of distinct coefficients in the vector $\bm{\Phi}_{tr.,t}$ which is obtained using the $t$th operator in \eqref{eq:20d}. Here, first, $\bm{\mathfrak{M}}_{4,t}\bm{\overline{\Phi}}$ is multiplied by $\bm{\overline{\Phi}}$, then, using a truncation function $\text{trunc}(\bm{\mathfrak{M}}_{4,t}\bm{\overline{\Phi}},Q)$, the first $Q$ elements of the vector $\bm{\mathfrak{M}}_{4,t}\bm{\overline{\Phi}}$ is truncated to get $\bm{\Phi}_{tr.,t}$. Its reverse operation, i.e., the conversion of $\bm{\Phi}_{tr.,t}$ to $\bm{\overline{\Phi}}$ can be performed using \eqref{eq:20e}, where $\text{merge}(\bm{a},\bm{b},Q)$ updates the first $Q$ elements of $\bm{b}$ by merging them with $\bm{a}$ of length $Q$.
	\end{definition}
	\begin{IEEEexample}
	    For the instance of the LC-DPD structure with $S=4$, $Q=4$, $r=1/2$, and $\nu =1$ in Fig.~\ref{fig:sys_mod}(b), $r^{-(g-1)}=(1/2)^{-(2-1)}=2$, thus, from \eqref{eq:seq_op}, $\bm{\mathfrak{M}}_{4} = \{\bm{\mathfrak{M}}_{4,1},\bm{\mathfrak{M}}_{4,2}\}$. As described earlier, the coefficients, $\overline{\phi}_{1,3}$, $\overline{\phi}_{1,4}$ are commonly shared with $\overline{\phi}_{1,1}$, $\overline{\phi}_{1,2}$ and $\overline{\phi}_{3,1}$, $\overline{\phi}_{3,2}$. From \eqref{eq:20d}, we can find the vectors, $\bm{\Phi}_{tr.,1}=[\overline{\phi}_{1,1},\overline{\phi}_{1,2},\overline{\phi}_{1,3}, \overline{\phi}_{1,4}]^T$ and $\bm{\Phi}_{tr.,2}=[\overline{\phi}_{3,1},\overline{\phi}_{3,2},\overline{\phi}_{1,3}, \overline{\phi}_{1,4}]^T$ from $\bm{\overline{\Phi}}=[\overline{\phi}_{1,1}, \overline{\phi}_{3,1}, \overline{\phi}_{1,2}, \overline{\phi}_{3,2}, \overline{\phi}_{1,3},\overline{\phi}_{1,4}]^T$ using the operators $\bm{\mathfrak{M}}_{4,1}$ and $\bm{\mathfrak{M}}_{4,2}$ as given by \eqref{eq:seq_op_t} which are:
	    	\begin{align}\label{eq:mat_ex_m4}\nonumber
	    	\arraycolsep=1.4pt\def\arraystretch{0.9}
	    	\bm{\mathfrak{M}}_{4,1}=\footnotesize \left[\begin{array}{@{}c|c@{}}
	    	\arraycolsep=1.4pt\def\arraystretch{0.9}
	    	\begin{matrix}
	    	1 & 0 & 0 & 0\\
	    	0 & 0 & 1 & 0\\
	    	0 & 0 & 0 & 0\\
	    	0 & 0 & 0 & 0
	    	\end{matrix}
	    	& 
	    	\arraycolsep=1.4pt\def\arraystretch{0.9}
	    	\begin{matrix}
	    	0 & 0 \\
	    	0 & 0 \\
	    	1 & 0 \\
	    	0 & 1
	    	\end{matrix} \\
	    	\hline
	    	\arraycolsep=1.4pt\def\arraystretch{0.9}
	    	\begin{matrix}
	    	0 & 1 & 0 & 0\\
	    	0 & 0 & 0 & 1
	    	\end{matrix}
	    	& 
	    	\arraycolsep=1.4pt\def\arraystretch{0.9}
	    	\begin{matrix}
	    	0 & 0 \\
	    	0 & 0 
	    	\end{matrix}
	    	\end{array}\right],\;\; \bm{\mathfrak{M}}_{4,2}=\footnotesize \left[\begin{array}{@{}c|c@{}}
	    	\arraycolsep=1.4pt\def\arraystretch{0.9}
	    	\begin{matrix}
	    	0 & 1 & 0 & 0\\
	    	0 & 0 & 0 & 1\\
	    	0 & 0 & 0 & 0\\
	    	0 & 0 & 0 & 0
	    	\end{matrix}
	    	& 
	    	\arraycolsep=1.4pt\def\arraystretch{0.9}
	    	\begin{matrix}
	    	0 & 0 \\
	    	0 & 0 \\
	    	1 & 0 \\
	    	0 & 1
	    	\end{matrix} \\
	    	\hline
	    	\arraycolsep=1.4pt\def\arraystretch{0.9}
	    	\begin{matrix}
	    	1 & 0 & 0 & 0\\
	    	0 & 0 & 1 & 0
	    	\end{matrix}
	    	& 
	    	\arraycolsep=1.4pt\def\arraystretch{0.9}
	    	\begin{matrix}
	    	0 & 0 \\
	    	0 & 0 
	    	\end{matrix}
	    	\end{array}\right].
	    	\end{align}
	    	Also, using this example, we can realize \eqref{eq:20e}.
	\end{IEEEexample}

	\begin{algorithm}[!t]
		{\small
			\caption{\small Estimation of coefficients for LC-DPD (Method-III).}\label{algo_4}
			\begin{algorithmic}[1]
				\Require The values of $\rho$, $\lambda_0$, $\mu$, $\bm{\mathfrak{M}}_1$, $\bm{\mathfrak{M}}_2$, $\bm{\mathfrak{M}}_{3,F}$, $\bm{\mathfrak{M}}_{3,B}$, $\bm{\tilde{\overline{\Phi}}}^{(0)}$, and $\mathcal{N}$
				\Ensure The estimated coefficients $\bm{\widehat{\overline{\Phi}}}$
				\State $\bm{P}_t^{(0)}=\text{diag}(\underbrace{\mu\bm{I}_{Q},\cdots,\mu\bm{I}_{Q}}_{T_{1}=Sr^{(\nu+g-1)}})$; $t\in\{1,\cdots,r^{-(g-1)}\}$
				\State $\bm{\xi}_t^{(0)}=\lambda_0\bm{I}_{T_{1}}$; $t\in\{1,\cdots,r^{-(g-1)}\}$
				\State Operate $\bm{\tilde{\Phi}}^{(0)} = \bm{\mathfrak{M}}_1\bm{\tilde{\overline{\Phi}}}^{(0)}$  and  assign $\bm{\Phi}^{(0)}=\bm{\tilde{\Phi}}^{(0)}$\label{algo4:step3} 
				\State Using obtained $\bm{X}(1)$ followed by $\bm{Y}(1)$ and then $\bm{\Psi}^{'}$, find $\bm{\Upsilon}(1)$, $\bm{\overline{\Psi^{'}}}(1) = \bm{\mathfrak{M}}_2 \bm{\Psi}^{'}(1)$, and $\bm{\overline{\Upsilon}}(1)$\label{algo4:step4}
				\State Compute $\bm{\tilde{X}}(1) = \bm{\Upsilon}(1)^T\bm{\tilde{\Phi}}^{(0)}$\label{algo4:step5}
				\State $n=0$
				\Repeat 
				\State $t=1$
				\Repeat
				\Repeat
				\State Operate $\bm{\tilde{\Phi}}^{(n)} = \bm{\mathfrak{M}}_1\bm{\tilde{\overline{\Phi}}}^{(n)}$  and  assign $\bm{\Phi}^{(n)}=\bm{\tilde{\Phi}}^{(n)}$\label{algo4:step11} 
				\State Using obtained $\bm{X}(n+1)$ followed by $\bm{Y}(n+1)$ and then $\bm{\Psi}^{'}(n+1)$, find $\bm{\Upsilon}(n+1)$, $\bm{\overline{\Psi^{'}}}(n+1) = \bm{\mathfrak{M}}_2 \bm{\Psi}^{'}(n+1)$, $\bm{\Psi}^{'}_t(n+1) = \text{trunc}(\bm{\mathfrak{M}}_{4,t}\bm{\overline{\Psi}}^{'}(n+1),Q )$, $\bm{\Upsilon}_t(n+1)$, and $\bm{\Phi}_{tr.,t}^{(n)} = \text{trunc}(\bm{\mathfrak{M}}_{4,t}\bm{\tilde{\overline{\Phi}}}^{(n)},Q)$\label{algo4:step12} 
				\State $\bm{\tilde{X}}(n+1) = \bm{\Upsilon}(n+1)^T\bm{\tilde{\Phi}}^{(n)}$\label{alog4:step13}
				\State $\bm{E}(n+1)=\bm{X}(n+1)-\bm{\tilde{X}}(n+1)$
				\State $\bm{\overline{E}}(n+1) = \bm{\mathfrak{M}}_2(\bm{E}(n+1)\bigotimes \bm{1}_Q)$
				\State $\bm{E}_t(n+1)=\bm{\mathfrak{M}}_{4,t}\bm{\overline{E}}(n+1)$
				\State $\bm{\xi}_t^{(n+1)}=\rho \bm{\xi}_t^{(n)}+\bm{I}_{_{T_{1}}}-\rho \bm{I}_{_{T_{1}}}$
				\State $\bm{Z}_t^{(n+1)} = \bm{\Upsilon}_{t}^T\hspace{-1mm}(n+1)\bm{P}_t^{(n)}\bm{\Upsilon}_{t}^{*}\hspace{-0.8mm}(n+1) + \bm{\xi}_t^{(n+1)}$\label{algo4:step18}
				\State $\bm{P}_t^{(n+1)} = (\bm{P}_t^{(n)}-\bm{P}_t^{(n)}\bm{\Upsilon}_{t}^{*}\hspace{-0.8mm}(n+1){\bm{Z}_t^{(n+1)}}^{-1}\bm{\Upsilon}_{t}^T\hspace{-1mm}(n+1)\bm{P}_t^{(n)}){\bm{\Xi}_t^{(n+1)}}^{-1}$\label{algo4:step19}	
				\State $\bm{\Phi}_{tr.,t}^{(n+1)} = \bm{\Phi}_{tr.,t}^{(n)} + (\bm{P}_t^{(n+1)}\bm{\Upsilon}_{t}^{*}\hspace{-0.8mm}(n+1)\bm{1}_{T_1})\bigodot \bm{E}_t(n+1)$\label{algo4:step20}
				\State $\bm{\tilde{\overline{\Phi}}}^{(n+1)} = \bm{\mathfrak{M}}_{4,t}^T\text{merge}(\bm{\Phi}_{tr.,t}^{(n+1)},\bm{\mathfrak{M}}_{4,t}\bm{\tilde{\overline{\Phi}}}^{(n)},Q)$
				\State $n=n+1$
				\Until{$n\;\%\;\mathcal{N} = = 0$}
				\State $t = t + 1$
				\Until{$t>r^{-(g-1)}$}
				\Until{$\bm{\tilde{\overline{\Phi}}}^{(n)}$ converges}
				\State $\bm{\widehat{\overline{\Phi}}}=\bm{\tilde{\overline{\Phi}}}^{(n)}$
			\end{algorithmic}
		}
	\end{algorithm}
	Now, use of the sequence of operators, $\bm{\mathfrak{M}}_{4}$ is described in Algorithm~\ref{algo_4} to correlate the common coefficients with the remaining distinct coefficients. Apart from the input parameters in Algorithm~\ref{algo_2}, $\bm{\mathfrak{M}}_{4}$ and the number $\mathcal{N}$ (defined later) are inputted in Algorithm~\ref{algo_4}. Then, it initializes the correlation matrix $\bm{P}_t$ and the forgetting matrix $\bm{\xi}_t$ for $t\in\{1,\cdots,r^{-(g-1)}\}$. In Steps \ref{algo4:step3}, \ref{algo4:step4}, and \ref{algo4:step5}, similar to earlier algorithms, it determines the initial values of $\bm{\tilde{X}}$ and $\bm{\overline{\Upsilon}}$. Thereafter, three nested loops are initialized. In Steps \ref{algo4:step11} and \ref{algo4:step12} of the innermost loop, the algorithm determines $\bm{\Upsilon}_{t}$ and $\bm{\Phi}_{tr.,t}$ using \eqref{eq:20d} in the $n$th iteration. Steps \ref{alog4:step13} to \ref{algo4:step20} follow the process to update $\bm{\Phi}_{tr.,t}$ in the current iteration. Then, $\bm{\Phi}_{tr.,t}$ is converted back to $\bm{\tilde{\overline{\Phi}}}$ using \eqref{eq:20e}. This process repeats for $\mathcal{N}$ iterations to correlate the common coefficients to the $t$th set of distinct coefficients. Thereafter, $t$ increases by unity to establish the correlation of the common coefficients with next set of distinct coefficients. Thus, the two inner loops repeat until $t>r^{-(g-1)}$ and at this point, the algorithm completes the one cycle to correlate the common coefficients with all sets of distinct coefficients. The outermost loop repeats this cycle until $\bm{\tilde{\overline{\Phi}}}$ converges. 
	
	\subsubsection*{Performance and Complexity}
	Algorithm~\ref{algo_4} performs better than Algorithms \ref{algo_2} and \ref{algo_3}, because, the common coefficients establish the correlation with all distinct coefficients in the vector $\bm{\overline{\Phi}}$. Therefore, using it, the generated predistorted signal vector $\bm{X}$ gives the better linearization of the PAs in the subarray. Moreover, for the fair comparison of this algorithm with the other earlier algorithms in terms of computational complexity, we assign $\mathcal{N}=1$. The complexity of a correlation cycle depends on the dominant matrix multiplications in Steps  \ref{algo4:step18}, \ref{algo4:step19}, and \ref{algo4:step20}. For the correlation of the common coefficients with the $r^{-(g-1)}$ distinct set of coefficients, the complexities of the three steps are: $O(Q^2T_1+QT_1)r^{-(g-1)}$, $O(2Q^3T_1+2Q^2T_1+QT_1)r^{-(g-1)}$, and $O(Q^2T_1+QT_1)r^{-(g-1)}$. Thus, considering the dominant term, the complexity is $O(Q^3T_1)r^{-(g-1)}$. Taking $r^{-(g-1)}$ inside the big O, the complexity can be approximated as: $O(Q^3T_1)r^{-(g-1)}\approx O(Q^3\overline{\sigma}_g)$. As $\overline{\sigma}_g>T_1$, but, $<S$, hence, the complexity of Algorithm~\ref{algo_4} is greater than Algorithms \ref{algo_3}, but, it is lesser than that of Algorithms \ref{algo_1} and \ref{algo_2}.
	
	\begin{figure*}[!t]
		\centering
		\subfigure[FF-DPD and Single-DPD]{\includegraphics[width=2.2in]{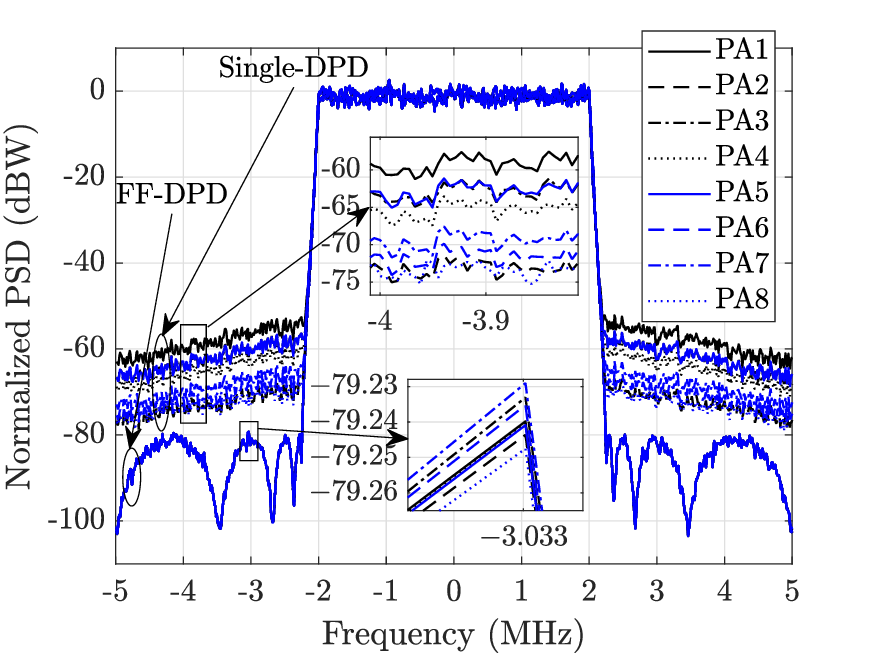}}
		\subfigure[LC-DPD I]{\includegraphics[width=2.2in]{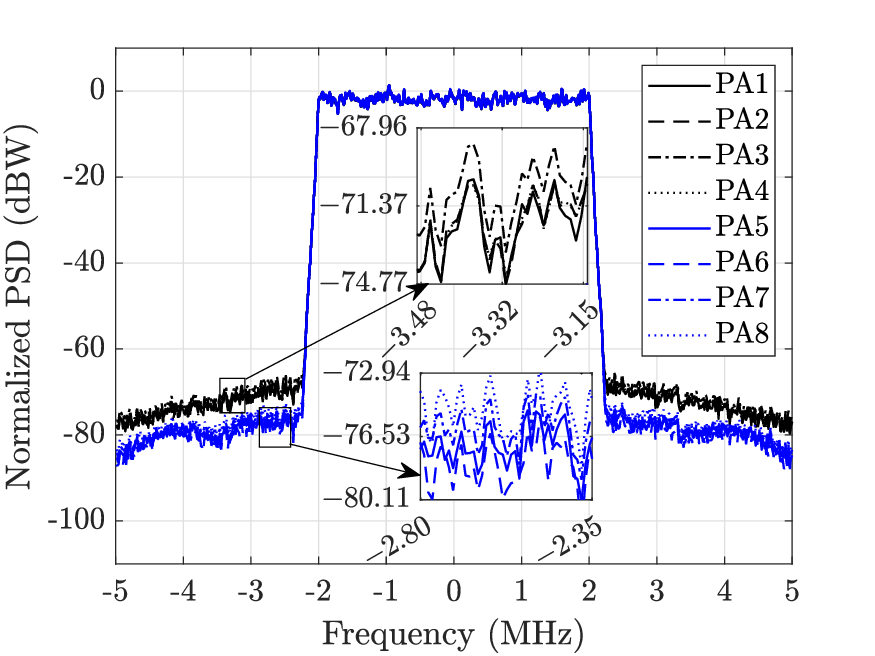}}
		\subfigure[LC-DPD II]{\includegraphics[width=2.2in]{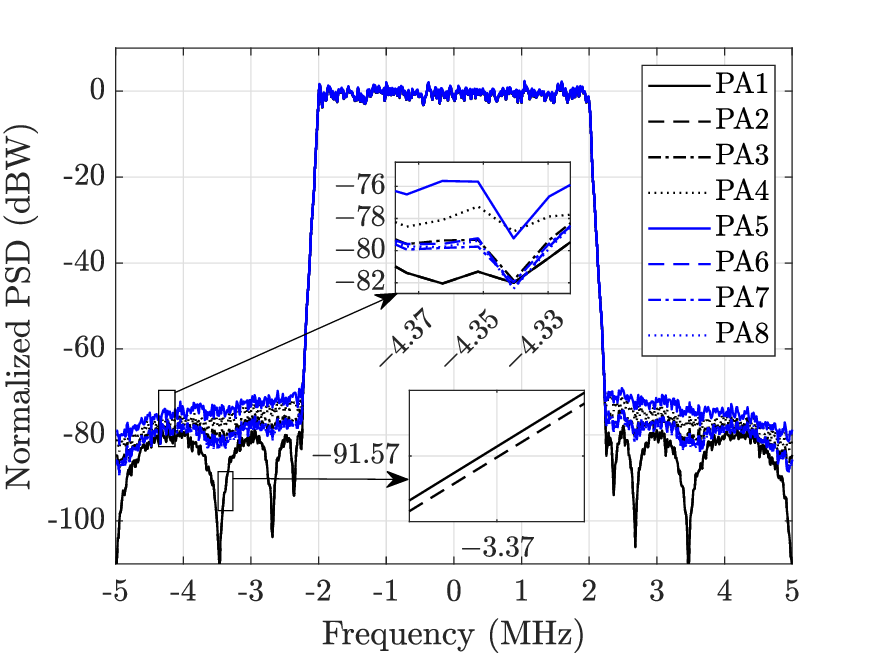}}
		\subfigure[LC-DPD III]{\includegraphics[width=2.2in]{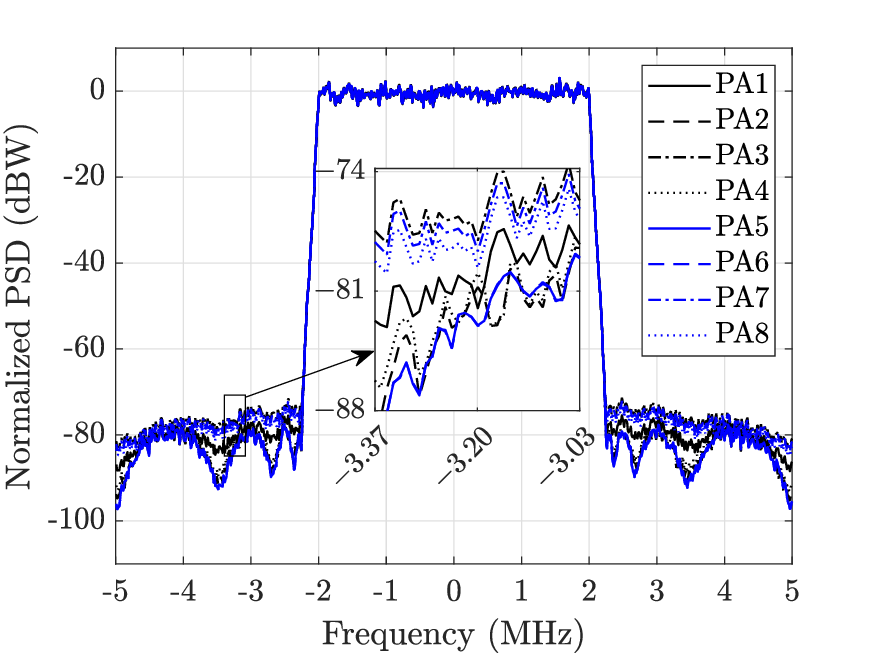}}
		\subfigure[EVM]{\includegraphics[width=4.48in]{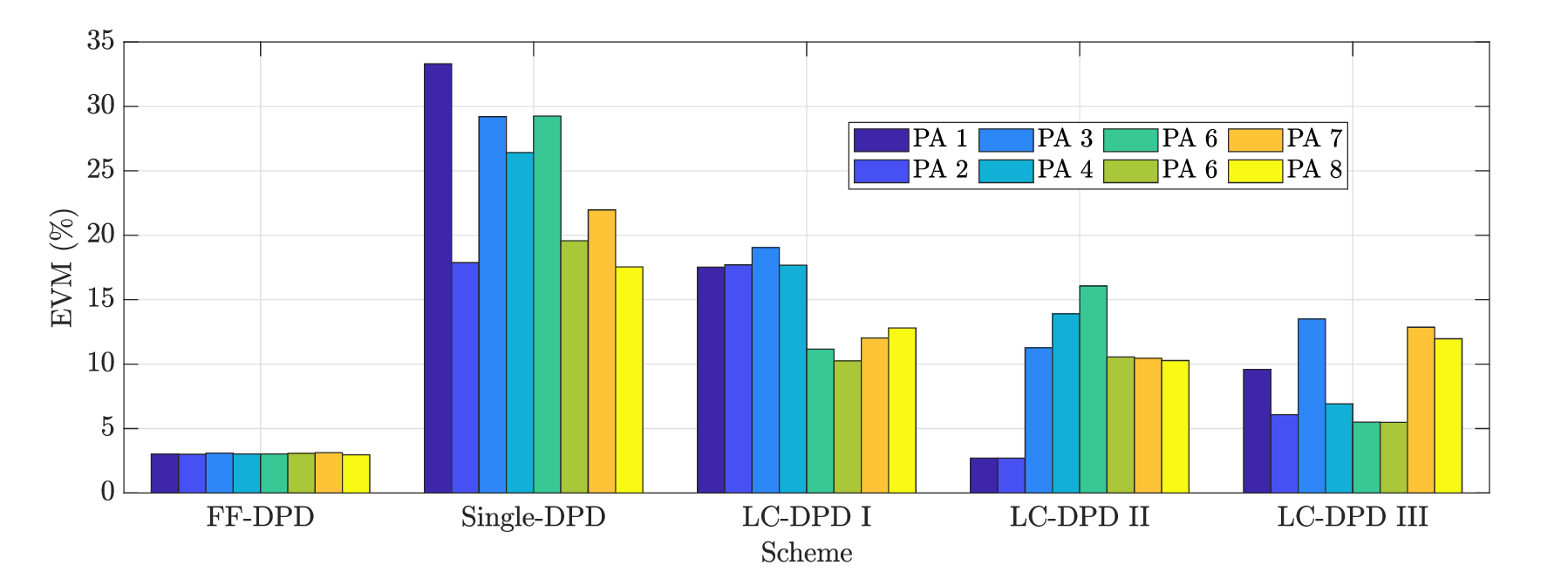}}
		\caption{\small Linearization of the PAs in a subarray using different schemes.}
		\label{fig:performance_comp}\vspace{-4mm}
	\end{figure*}
	
   \section{Numerical Results}\label{sec:num_rel}
   
   \subsection{Evaluation Environment}
	To evaluate the performance of the proposed analysis, we consider a subarray of $S=8$ PAs. The PAs follow Saleh model as in~\cite{liu49}. The parameters, $\alpha_{a,i}$ and $\beta_{a,i}$ for the AM/AM distortion and $\alpha_{\phi,i}$ and $\beta_{\phi, i}$ for the AM/PM distortion are given by: $\alpha_{a,i}=0.9445+0.1 u_{a,i}$, $\beta_{a,i}=0.5138+0.1 v_{a,i}$, $\alpha_{\phi, i}=4.0033+ u_{\phi, i}$, and $\beta_{\phi, i}=9.1040+ v_{\phi, i}$, where $u_{a,i}$, $v_{a,i}$, $u_{\phi, i}$ and $v_{\phi, i}$ are uniformly distributed over $[0,1]$ for $i\in\{1,\cdots,S\}$. The GMP used for a DPD has the order $P=5$ and each order has the memory length, $M=5$. But, as described using Fig.~\ref{fig:coeff_val}, the BFs with indices in the set $\mathcal{I}$ have the nonzero coefficients; thus $Q=10$. Further, for the LC-DPD scheme, the BFs are arranged in decreasing order of their dominance. The arrangement is represented using the indices of the BFs as: $\{4,5,14,15,19,20,24,25,9,10\}$. Moreover, for the LC-DPD structure, the geometric sequence in~\eqref{eq:seq_grp} has the following parameters' values: $g=2$, $n_1=4$ $n_2=6$, $\nu = 1$, and $r=1/2$. The bandwidth of the input signal $s(n)$ is $4$ MHz. To get the insights on the linearization using the obtained results, the in-band average powers of the power spectral density (PSD) of the input signal $s(n)$ and the output of the PAs, $y_l(n)$; $l\in \{1\cdots S\}$, are normalized to $0$ dB. Moreover, for the algorithms based on ILA-RPEM, the input parameters are set as: $\lambda_0=0.99$,  $\mu = 0.2$, and $\rho = 0.95$. The linearization of the PAs is determined using the error vector magnitudes (EVMs) of their outputs with respect to the reference message signal $s(n)$. It is computed as: $\text{EVM}_{PA_l}=\sqrt{\sum_{n}(y_l(n)-s(n))^2/\sum_{n}s^2(n)}$; $l\in \{1,\cdots,S\}$. The simulation for the proposed analysis is performed using MATLAB/Simulink${}^\text{\textregistered}$. 
	
	\subsection{Performance Comparison}
	In Fig.~\ref{fig:performance_comp}, the different DPD schemes are compared in terms of linearization of the PAs in the subarray. It can be observed in Fig.~\ref{fig:performance_comp}(a) that FF-DPD gives the best performance to linearize all the PAs, whereas, single-DPD has the least performance. Because, in FF-DPD, each PA has a separate DPD to linearize itself, but, in single-DPD, all the PAs are linearized using a single DPD. Also, from the bar plot in Fig.~\ref{fig:performance_comp}(e), all the PAs have almost same EVMs and their values are around $3.04\%$. Whereas, for single-DPD, the EVMs values of the PAs are different and the maximum value goes upto $33.31\%$. If we compare the three LC-DPD schemes in Figs. \ref{fig:performance_comp}(b), \ref{fig:performance_comp}(c), and \ref{fig:performance_comp}(d), the linearization performance of LC-DPD~I is least, because, LC-DPD coefficients in it are determined using the structure of FF-DPD where correlation of the common coefficients are least to the distinct coefficients. Therefore, although, it is giving the better performance than single-DPD, but, none of the PAs is linearized properly. The average EVMs for the first four and the next four PAs are $17.99\%$ and  $11.56\%$, respectively. In LC-DPD II scheme, the structure of LC-DPD is completely exploits, but, the common coefficients are correlated with distinct coefficients for the first two PAs. Therefore, the linearization of these PAs are same as FF-DPD with average EVM equal to $2.71\%$, while the linerization of the remaining $6$ PAs is less and their average EVM value is $12.09\%$. In LC-DPD~III, the common coefficients are partially correlated with each of the distinct coefficients, therefore, its overall performance is better than the previous two schemes. The average EVM values of LC-DPD~I, LC-DPD~II, and LC-DPD~III schemes are $14.77\%$, $9.74\%$, and $8.98\%$, respectively.
	
	\section{Conclusion}
	In this work, we have proposed two schemes, FF-DPD and LC-DPD to fully linearize the PAs in a subarray of a mMIMO transmitter. Although, FF-DPD provides the best performance but it has high complexity. Using the structure of FF-DPD, we derive a less complex LC-DPD. For the two schemes, four algorithms based on ILA-RPEM are described and their performances and complexities are investigated. From the obtained results we find that FF-DPD almost linearizes the PAs fully with on average $3.04\%$ EVM. The computational complexities of the three algorithms for LC-DPD is less complex, but, their performances in EVM are less than the algorithm for FF-DPD. Furthermore, among the three algorithms for LC-DPD, the third algorithm (LC-DPD III) provides the better performance as it better correlates the common coefficients to the distinct coefficients of the scheme.

	\bibliographystyle{IEEEtran}
	\bibliography{references_PA_DPD}
	
\end{document}